\newcommand \Mpc {h^{-1}{\rm Mpc}}
\newcommand \Mpcx {{\rm Mpc}}
\newcommand \kpc {h^{-1}{\rm kpc}}

\newcommand \farcm{\hbox{$.\!\!^{\prime}$}}
\newcommand \arcm{\hbox{$^{\prime}$}}
\newcommand \arcs{\hbox{$^{\prime\prime}$}}

\newcommand \kms {{\rm km~s}^{-1}}
\newcommand \msun {h^{-1} M_\odot}
\newcommand \beqn {\begin{equation}}
\newcommand \eeqn {\end{equation}}
\newcommand \nczhecsnew {22,680 }

\newcommand \nczmem {10,145 }

\documentclass[apj]{emulateapj}
\usepackage{epsfig}
\usepackage{natbib}
\usepackage{url}
\newcommand{\noprint}[1]{}

\begin{document}

\title{Measuring the Ultimate Halo Mass of Galaxy Clusters: Redshifts and Mass Profiles from the Hectospec Cluster Survey (HeCS)}
\shorttitle{HeCS: Ultimate Mass of Clusters}
\shortauthors{Rines et al.}

\author{Kenneth Rines\altaffilmark{1,2}, Margaret J. Geller\altaffilmark{2}, 
Antonaldo Diaferio\altaffilmark{3,4}, and Michael J. Kurtz\altaffilmark{2}} 
\email{kenneth.rines@wwu.edu}

\altaffiltext{1}{Department of Physics \& Astronomy, Western Washington University, Bellingham, WA 98225; kenneth.rines@wwu.edu}
\altaffiltext{2}{Smithsonian Astrophysical Observatory, 60 Garden St, MS 20, Cambridge, MA 02138}
\altaffiltext{3}{Universit\`a di Torino,
Dipartimento di Fisica, Torino, Italy; diaferio@ph.unito.it}
\altaffiltext{4}{Istituto Nazionale di Fisica Nucleare (INFN) -- Sezione di Torino, Italy}

\begin{abstract}

The infall regions of galaxy clusters represent the largest gravitationally 
bound structures in a $\Lambda$CDM universe.  Measuring cluster
mass profiles into the infall regions provides an estimate of the ultimate 
mass of these haloes.  We use the caustic technique to measure cluster
mass profiles from galaxy redshifts obtained with the Hectospec Cluster 
Survey (HeCS), an extensive spectroscopic survey of 
galaxy clusters with MMT/Hectospec.  
We survey 58 clusters selected by X-ray flux at
0.1$<$$z$$<$0.3.  
The survey includes \nczhecsnew unique MMT/Hectospec redshifts for 
individual galaxies; \nczmem of these galaxies are cluster members.  
For each cluster we acquired high signal-to-noise spectra for $\sim$200 
cluster members and  a comparable number of foreground/background 
galaxies.  The cluster members trace out infall patterns around the clusters.  
The members define a very narrow red sequence.  We demonstrate that the 
determination of velocity dispersion is insensitive to the inclusion of 
bluer members (a small fraction of  the cluster population).  We apply the 
caustic technique to define membership and estimate the mass profiles to large
radii.  The ultimate halo mass of clusters 
(the mass that remains bound in the far future of a $\Lambda$CDM universe) 
is on average (1.99$\pm$0.11)$M_{200}$, a new observational 
cosmological test in essential agreement with 
simulations. Summed profiles binned in $M_{200}$ and in $L_X$ demonstrate 
that the predicted NFW form of the density profile is a remarkably good 
representation of the data in agreement with weak lensing results extending 
to large radius. The concentration of these summed profiles is also consistent 
with theoretical predictions.

\end{abstract}


\section{Introduction}

Clusters of galaxies are the most massive virialized
systems in the universe.  Clusters
are surrounded by infall regions where galaxies are bound to the
cluster but are not in dynamical equilibrium in the cluster potential.  If dark energy behaves like a
cosmological constant, cluster infall regions are the largest
gravitationally bound structures in the universe.  Thus, measuring 
cluster mass profiles at large radii provides an estimate of the 
ultimate halo mass of these systems \citep[][]{nl02,busha05,dunner06}.

The Cluster and Infall Region Nearby Survey (CAIRNS)
pioneered the detailed observational study of cluster infall regions.  
CAIRNS studied nine nearby galaxy clusters and their
infall regions with extensive spectroscopy \citep{cairnsi,cairnsha}
and near-infrared photometry from the Two-Micron All-Sky Survey
\citep[][]{cairnsii}.  The nine CAIRNS clusters display a
characteristic trumpet-shaped pattern in radius-redshift phase space
diagrams.  These patterns were first predicted for simple spherical
infall onto clusters \citep{kais87,rg89}, but later work showed that
these patterns reflect the dynamics of the infall region
\citep[][hereafter DG and D99]{dg97,diaferio1999}.
The Cluster Infall Regions in the Sloan Digital Sky Survey
\citep[][hereafter CIRS]{cirsi} project extended this analysis to 
72 X-ray-selected clusters in the Sloan Digital Sky Survey
\citep[SDSS,][]{sdss}.  CIRS showed that these infall patterns are
ubiquitous in nearby X-ray clusters.

Using numerical simulations, DG and D99 showed that the amplitude of
the caustics is a measure of the escape velocity from the cluster;
identification of the caustics therefore allows a determination of the
mass profile of the cluster on scales $\lesssim 10\Mpc$.  In
particular, nonparametric measurements of caustics yield cluster mass
profiles accurate to $\sim$50\% on scales $\lesssim 10 h^{-1}$ Mpc
when applied to Coma-size clusters extracted from cosmological
simulations. 
\citet{serra11} confirm these results for clusters across a broader 
mass range and they show that the dominant source of 
uncertainty in individual cluster mass profiles is projection effects
from departures from spherical symmetry.  
The caustic technique
assumes only that galaxies trace the velocity field. Indeed,
simulations suggest that little or no velocity bias exists on linear
and mildly non-linear scales
\citep{kauffmann1999a,kauffmann1999b,diemand04,faltenbacher05}.  This
conclusion is supported observationally by the excellent agreement
between the cluster virial mass function and other cosmological
probes \citep{rines08}.

CAIRNS and CIRS showed that caustic masses of clusters agree well with
mass estimates from both X-ray observations and Jeans' analysis at
small radii \citep[][CIRS]{cairnsi}.  \citet{lokas03} confirm that the
mass of Coma estimated from higher moments of the velocity
distribution agrees well with the caustic mass estimate \citep{gdk99}.

The caustic technique provides an estimate of the mass profile of 
clusters.  For instance, CAIRNS and CIRS showed that cluster mass 
profiles fall off more steeply at large radii than an isothermal sphere. 
Thus, caustic mass profiles probe the structure of dark matter haloes,
and these profiles can be compared to those determined from 
gravitational lensing \citep[e.g.,][]{umetsu11b,geller12b}.

At large radii, neither galaxies nor intracluster gas should
be in equilibrium, thus invalidating the use of the virial theorem or 
hydrostatic equilibrium at these radii.  The caustic technique and 
gravitational lensing are the only cluster mass estimators that do 
not rely on the equilibrium assumption.  Gravitational lensing 
is necessarily contaminated by line-of-sight structure unrelated to the 
cluster; this contamination becomes larger at larger radii \citep[e.g.,][]{hoekstra11}.
Despite this potential difficulty,
\citet{diaferio05} showed that caustic masses agree with weak
lensing masses in three clusters at moderate redshift.  

CAIRNS and CIRS showed that infall patterns are well defined in
observations of nearby massive clusters.  In fact, the infall patterns
or ``caustics'' have significantly higher contrast in the CAIRNS
observations than in the simulations of DG and D99.  The CAIRNS and
CIRS clusters are fairly massive clusters and generally have
little surrounding large-scale structure \citep[but
see][]{rines01b,rines02}.  

One might suspect that the presence of
infall patterns is limited to massive, isolated clusters.  However,
other investigators have found infall patterns around the Fornax
Cluster \citep{drink}, the Shapley Supercluster
\citep{rqcm}, an ensemble cluster comprised of poor clusters in the
Two Degree Field Galaxy Redshift Survey \citep{bg03}, and even the
galaxy group associated with NGC 5846 \citep{mahdavi05}.  Caustics 
are also identifiable in X-ray groups in SDSS \citep{4dlx}.

The presence of caustics in all nine CAIRNS clusters and in all 72 CIRS
clusters suggests that they are ubiquitous in nearby, massive
clusters.  Because clusters and especially cluster infall regions form
quite late in the evolution of structure, infall
patterns evolve even at modest redshifts \citep[see an animation in ][]{geller11}.  \citet{diaferio05}
showed that infall patterns exist in three clusters at moderate
redshift, although these clusters were not carefully selected.
Similarly, \citet{lemze09} showed that the caustic mass profile 
of A1689 ($z$=0.18) agrees with the mass profile determined from a joint 
analysis of X-ray and lensing data.

Inspired by these results, we conducted a systematic survey of
infall regions at $z$$\approx$0.2.  The Hectospec
Cluster Survey (HeCS), includes
MMT/Hectospec spectra of large samples of galaxies in the infall regions of X-ray-selected
clusters at $z$$=$0.1-0.3.  HeCS is
the first systematic spectroscopic survey of cluster infall regions at
$z$$\gtrsim$0.1.

Spectroscopic data for clusters at moderate redshift can test the
robustness of scaling relations between different cluster observables
that correlate with cluster mass.  For instance, the integrated
Sunyaev-Zeldovich decrement $Y_{SZ}$ has small scatter in simulations
\citep{hallman06}, as does the X-ray observable $Y_X=M_{gas}T_X$
\citep{kravtsov06}.  The $Y_X$ parameter also seems to have low
scatter in observations, although the estimated masses and $Y_X$ are
derived from the same data \citep{nagai07}.  \citet{hecsysz} studied 
the relation between $Y_{SZ}$ and the dynamics of galaxies in a 
sample of 15 clusters.  The dynamical properties correlate strongly 
with increasing $Y_{SZ}$ but with significant scatter.

We describe the data and the cluster sample in $\S$ 2.  In $\S$ 3, we
review the caustic technique, use it to estimate the cluster mass
profiles, and estimate the ultimate halo masses of clusters.  To mitigate systematic 
uncertainties due to projection effects, we analyze ensemble 
clusters in $\S$ 4 and compare the ensemble mass profiles with the 
theoretical NFW profile \citep{nfw97} and with profiles
determined from gravitational lensing. We discuss the color 
distribution of cluster members in $\S$ 5 and show that our red-sequence
selection does not bias our dynamical measurements.   We discuss our results and
conclude in $\S 6$.  We assume $H_0 = 100 h~\kms, \Omega _m = 0.3,
\Omega _\Lambda = 0.7$ throughout.  We measure cluster masses $M_\Delta$, 
defined as the mass enclosed within the radius $r_\Delta$ that encloses an 
average density of $\Delta \rho_c(z)$, where $\rho_c(z) = 3H_0^2E^2(z)/8\pi G$ 
is the critical density at redshift $z$ and $E^2(z)=\Omega_m(1+z)^3+\Omega_\Lambda+(1-\Omega_m-\Omega_\Lambda)(1+z)^2$.

\section{The HeCS Cluster Sample}

We construct the HeCS sample to take advantage of two large-area
public surveys: the Sloan Digital Sky Survey \citep[SDSS,][]{sdss} and
the ROSAT All-Sky Survey \citep[RASS,][]{rass}.  In particular, we
utilize existing X-ray cluster catalogs based on RASS data to define a
flux-limited cluster sample.  We then match these systems to the
imaging footprint of the SDSS Data Release 6 \citep[DR6,][]{dr6}.  The
accurate SDSS multicolor photometry enables 
selection of candidate cluster members using the red-sequence technique
\citep[e.g.,][]{gladders00}.

We obtained MMT/Hectospec spectroscopy of 400-550 candidate
members per cluster (for three clusters, we obtained more than 550 spectra).  
Hectospec redshifts enable robust membership
classification and estimates of the virial masses of the
clusters.  The wide field-of-view of Hectospec allows us to
simultaneously probe the virial and infall regions of these clusters.

The earlier CIRS project used spectroscopic data from the Sloan
Digital Sky Survey \citep[SDSS,][]{sdss} to study clusters at
$z$$\leq$0.1 identified in X-ray cluster catalogs based on the ROSAT
All-Sky Survey.  At $z$$>$0.1, the SDSS redshift survey is not dense enough 
for accurate assessment of cluster masses. 

\subsection{Sloan Digital Sky Survey \label{sdssdesc}}

The Sloan Digital Sky Survey \citep[SDSS,][]{sdss} is a wide-area
photometric and spectroscopic survey at high Galactic latitudes.  The
Sixth Data Release (DR6) of SDSS includes 8417 square degrees of
imaging data \citep{dr6}.

From a comparison of SDSS with the Millennium Galaxy Catalogue,
\citet{2004MNRAS.349..576C} conclude that there is a photometric
incompleteness of $\sim$7\% due to galaxies misclassified as stars or
otherwise missed by the SDSS photometric pipeline.  For our purposes,
the incompleteness is not important provided sufficient numbers of
cluster galaxies do have spectra.  Further, the cluster galaxies we
focus on here are intrinsically luminous and have higher surface
brightnesses than less luminous galaxies.  Thus, the photometric
incompleteness is probably somewhat smaller for these galaxies.
 
The spectroscopic limit of the main galaxy sample of SDSS is $r$=17.77
after correcting for Galactic extinction \citep{strauss02}.  CAIRNS
and CIRS show that infall patterns are easily detectable in clusters
sampled to about $M^*+1$.  For SDSS, this limit is reached at $z\sim$0.1; 
deeper spectroscopic surveys are required to study more distant clusters.
Studying cluster infall regions at $z$$\sim$0.2 requires spectroscopy 
to a limit of $r\sim 21$.

\subsection{X-ray Cluster Surveys \label{xcs}}

We select the HeCS clusters from X-ray catalogs based on the ROSAT
All-Sky Survey \citep[RASS,][]{rass}.  RASS is a shallow
survey, but it is sufficiently deep to include nearby, massive
clusters.  RASS covers virtually the entire sky and is thus the most
complete X-ray cluster survey for nearby clusters.  The flux limits of
RASS-based surveys are $\approx$3$\times 10^{-12}$erg s$^{-1}$cm$^{-2}$ 
in the ROSAT band \citep{bcs,noras,2004A&A...425..367B}.

We restrict our study to systems with 0.10$\leq$$z$$\leq$0.30.  The
flux-limited HeCS sample consists of the 53 clusters from the 
Bright Cluster Survey \citep[BCS;][]{bcs} and ROSAT-ESO FLux-LimitEd 
X-ray cluster survey \citep[REFLEX;][]{2004A&A...425..367B} catalogs 
within the SDSS DR6 photometric footprint and with
$f_X$$\geq$5$\times 10^{-12}$erg s$^{-1}$cm$^{-2}$.  Due to scheduling
constraints on observing time, HeCS includes four additional clusters (A2187, A2396, A2631, 
and A2645) with $f_X$$\geq$3$\times 10^{-12}$erg s$^{-1}$cm$^{-2}$: 
A2645 is in REFLEX and the others are from the extended 
BCS \citep[eBCS;][]{ebcs}.  BCS splits A1758 into two components, 
neither  of which alone would lie above our flux limit; however, the NORAS 
catalog \citep{noras} merges these components, yielding a flux above 
our flux limit.  We also include an X-ray cluster, A750, that lies in the 
foreground  of MS0906+1110 (see $\S$\ref{individual} for details).
HeCS thus contains 58 clusters: 53 clusters in the flux-limited sample, four 
clusters with slightly smaller X-ray fluxes, and A750 (we do not count the 
subclusters of A1758 as individual clusters). 

A689 has a large contribution from a central BL Lac; \citet{giles11}
estimate that only 10\% of the BCS luminosity is from the intracluster medium. 
The long arrow in Figure \ref{hecslxz} shows that this cluster would not lie in the 
flux-limited sample with the corrected luminosity.  Similarly, the luminosity of 
A1758N (the brighter of A1758N/S) alone would remove it from the flux-limited 
sample (short arrow in Figure \ref{hecslxz}). 

Table \ref{sample} describes the basic properties of the HeCS
clusters.  X-ray luminosities are in the rest-frame ROSAT band and 
are from the BCS and REFLEX catalogs. 
Figure \ref{hecslxz} shows the (rest-frame) X-ray luminosities of the HeCS clusters compared
to the HiFluGCS survey \citep{hiflugcs} and to our previous survey CIRS.  HeCS contains systems
with significantly larger $L_X$ than CIRS, although there is a
substantial overlap at intermediate $L_X$, especially at $z$=0.10-0.15.  
The larger redshift limit of HeCS relative to CIRS probes
a much larger volume: the HeCS volume
is $\sim$10$^{8}$$h^{-3}\mbox{Mpc}^{3}$, a factor of $>$10 larger than the volume
probed by CIRS.  The number density of clusters declines with $L_X$
\citep[e.g.,][]{bohringer02}; a larger survey volume increases the sample of intrinsically more
luminous systems.

\begin{table*}[th] \footnotesize
\begin{center}
\caption{\label{sample} \sc HeCS Basic Properties}
\begin{tabular}{lcccrrcr}
\tableline
\tableline
\tablewidth{0pt}
Name &\multicolumn{2}{c}{X-ray Coordinates} & $z_\odot$ & $L_X /10^{43}$  & Catalog & $\sigma_p$ & $N_{m}$ \\ 
 & RA (J2000) & DEC (J2000) &   & erg~s$^{-1}$& & $\kms$ & \\ 
\tableline
       A267 & 28.1762 & 01.0125 & 0.2291 & 4.16 &  BCS & $ 972 ^{+63} _{-53}$ & 226 \\ 
     Zw1478 & 119.9190 & 53.9990 & 0.1027 & 0.64 &  BCS & $ 479 ^{+66} _{-46}$ & 82 \\ 
       A646 & 125.5470 & 47.1000 & 0.1273 & 1.22 &  BCS & $ 653 ^{+66} _{-51}$ & 264 \\ 
       A655 & 126.3610 & 47.1320 & 0.1271 & 1.91 &  BCS & $ 777 ^{+58} _{-47}$ & 315 \\ 
       A667 & 127.0190 & 44.7640 & 0.1452 & 1.32 &  BCS & $ 645 ^{+80} _{-58}$ & 148 \\ 
       A689 & 129.3560 & 14.9830 & 0.2789 & 9.62 &  BCS & $ 589 ^{+91} _{-62}$ & 163 \\ 
       A697 & 130.7362 & 36.3625 & 0.2812 & 5.15 &  BCS & $ 1002 ^{+97} _{-75}$ & 185 \\ 
       A750 & 137.2469 & 11.0444 & 0.1640 & -- &  BCS* & $ 681 ^{+56} _{-45}$ & 225 \\ 
     MS0906 & 137.2832 & 10.9925 & 0.1767 & 3.23 &  BCS & $ 664 ^{+87} _{-62}$ & 101 \\ 
       A773 & 139.4624 & 51.7248 & 0.2173 & 3.98 &  BCS & $ 1110 ^{+86} _{-70}$ & 173 \\ 
       A795 & 141.0240 & 14.1680 & 0.1374 & 1.65 &  BCS & $ 778 ^{+61} _{-50}$ & 179 \\ 
     Zw2701 & 148.1980 & 51.8910 & 0.2160 & 3.30 &  BCS & $ 652 ^{+74} _{-55}$ & 93 \\ 
       A963 & 154.2600 & 39.0484 & 0.2041 & 3.07 &  BCS & $ 956 ^{+80} _{-64}$ & 211 \\ 
       A980 & 155.6275 & 50.1017 & 0.1555 & 2.06 &  BCS & $ 1033 ^{+72} _{-59}$ & 222 \\ 
     Zw3146 & 155.9117 & 04.1865 & 0.2894 & 8.38 &  BCS & $ 858 ^{+103} _{-75}$ & 106 \\ 
       A990 & 155.9120 & 49.1450 & 0.1416 & 2.13 &  BCS & $ 655 ^{+82} _{-60}$ & 91 \\ 
     Zw3179 & 156.4840 & 12.6910 & 0.1422 & 1.34 &  BCS & $ 541 ^{+122} _{-73}$ & 69 \\ 
      A1033 & 157.9320 & 35.0580 & 0.1220 & 1.35 &  BCS & $ 677 ^{+55} _{-44}$ & 191 \\ 
      A1068 & 160.1870 & 39.9510 & 0.1386 & 2.18 &  BCS & $ 1028 ^{+106} _{-81}$ & 129 \\ 
      A1132 & 164.6160 & 56.7820 & 0.1351 & 1.92 &  BCS & $ 749 ^{+80} _{-61}$ & 160 \\ 
      A1201 & 168.2287 & 13.4448 & 0.1671 & 1.79 &  BCS & $ 683 ^{+68} _{-53}$ & 165 \\ 
      A1204 & 168.3324 & 17.5937 & 0.1706 & 2.13 &  BCS & $ 532 ^{+62} _{-46}$ & 92 \\ 
      A1235 & 170.8040 & 19.6160 & 0.1030 & 0.65 &  BCS & $ 584 ^{+62} _{-47}$ & 131 \\ 
      A1246 & 170.9912 & 21.4903 & 0.1921 & 2.31 &  BCS & $ 906 ^{+70} _{-57}$ & 226 \\ 
      A1302 & 173.3070 & 66.3990 & 0.1152 & 0.84 &  BCS & $ 650 ^{+62} _{-48}$ & 162 \\ 
      A1361 & 175.9170 & 46.3740 & 0.1159 & 0.99 &  BCS & $ 512 ^{+64} _{-47}$ & 195 \\ 
      A1366 & 176.2020 & 67.4130 & 0.1160 & 1.08 &  BCS & $ 616 ^{+62} _{-48}$ & 200 \\ 
      A1413 & 178.8260 & 23.4080 & 0.1412 & 3.71 &  BCS & $ 856 ^{+90} _{-68}$ & 116 \\ 
      A1423 & 179.3420 & 33.6320 & 0.2142 & 3.07 &  BCS & $ 759 ^{+64} _{-51}$ & 230 \\ 
      A1437 & 180.1040 & 03.3490 & 0.1333 & 2.12 &  BCS & $ 1233 ^{+102} _{-81}$ & 194 \\ 
      A1553 & 187.6959 & 10.5606 & 0.1668 & 2.17 &  BCS & $ 867 ^{+62} _{-51}$ & 171 \\ 
      A1682 & 196.7278 & 46.5560 & 0.2272 & 3.48 &  BCS & $ 996 ^{+80} _{-65}$ & 151 \\ 
      A1689 & 197.8750 & -1.3353 & 0.1842 & 7.06 &  REF & $ 1197 ^{+78} _{-65}$ & 210 \\ 
      A1758 & 203.1796 & 50.5496 & 0.2760 & 5.82 &  BCS & $ 674 ^{+99} _{-69}$ & 143 \\ 
      A1763 & 203.8257 & 40.9970 & 0.2312 & 4.72 &  BCS & $ 1261 ^{+81} _{-68}$ & 237 \\ 
      A1835 & 210.2595 & 02.8801 & 0.2506 & 11.77 &  BCS & $ 1151 ^{+80} _{-66}$ & 219 \\ 
      A1902 & 215.4226 & 37.2958 & 0.1623 & 1.67 &  BCS & $ 784 ^{+71} _{-56}$ & 130 \\ 
      A1918 & 216.3420 & 63.1830 & 0.1388 & 1.17 &  BCS & $ 545 ^{+76} _{-54}$ & 80 \\ 
      A1914 & 216.5068 & 37.8271 & 0.1660 & 5.03 &  BCS & $ 798 ^{+53} _{-44}$ & 255 \\ 
      A1930 & 218.1200 & 31.6330 & 0.1308 & 1.15 &  BCS & $ 577 ^{+75} _{-54}$ & 76 \\ 
      A1978 & 222.7750 & 14.6110 & 0.1459 & 1.29 &  BCS & $ 404 ^{+95} _{-56}$ & 63 \\ 
      A2009 & 225.0850 & 21.3620 & 0.1522 & 2.60 &  BCS & $ 715 ^{+57} _{-46}$ & 195 \\ 
    RXJ1504 & 226.0321 & -2.8050 & 0.2168 & 14.12 &  REF & $ 779 ^{+105} _{-75}$ & 120 \\ 
      A2034 & 227.5450 & 33.5060 & 0.1132 & 1.92 &  BCS & $ 942 ^{+64} _{-53}$ & 182 \\ 
      A2050 & 229.0680 & 00.0890 & 0.1191 & 1.18 &  BCS & $ 869 ^{+77} _{-61}$ & 106 \\ 
      A2055 & 229.6720 & 06.2110 & 0.1023 & 1.34 &  BCS & $ 676 ^{+90} _{-64}$ & 230 \\ 
      A2069 & 231.0410 & 29.9210 & 0.1139 & 2.46 &  BCS & $ 994 ^{+61} _{-52}$ & 441 \\ 
      A2111 & 234.9337 & 34.4156 & 0.2291 & 3.35 &  BCS & $ 741 ^{+65} _{-52}$ & 208 \\ 
      A2187 & 246.0591 & 41.2383 & 0.1829 & 1.56 &  eBCS & $ 631 ^{+83} _{-59}$ & 103 \\ 
      A2219 & 250.0892 & 46.7058 & 0.2257 & 6.10 &  BCS & $ 1151 ^{+63} _{-54}$ & 461 \\ 
     Zw8197 & 259.5480 & 56.6710 & 0.1132 & 0.80 &  BCS & $ 597 ^{+73} _{-53}$ & 76 \\ 
      A2259 & 260.0370 & 27.6702 & 0.1605 & 1.85 &  BCS & $ 855 ^{+76} _{-60}$ & 165 \\ 
    RXJ1720 & 260.0370 & 26.6350 & 0.1604 & 4.47 &  BCS & $ 860 ^{+40} _{-35}$ & 376 \\ 
      A2261 & 260.6129 & 32.1338 & 0.2242 & 5.55 &  BCS & $ 780 ^{+78} _{-60}$ & 209 \\ 
    RXJ2129 & 322.4186 & 00.0973 & 0.2339 & 5.65 &  BCS & $ 858 ^{+71} _{-57}$ & 325 \\ 
      A2396 & 328.9198 & 12.5336 & 0.1919 & 1.86 &  eBCS & $ 935 ^{+90} _{-70}$ & 176 \\ 
      A2631 & 354.4206 & 00.2760 & 0.2765 & 4.15 &  eBCS & $ 851 ^{+96} _{-72}$ & 173 \\ 
      A2645 & 355.3200 & -9.0275 & 0.2509 & 2.85 &  REF & $ 549 ^{+78} _{-55}$ & 61 \\ 
\tableline
\end{tabular}
\end{center}
\end{table*}

\subsection{MMT/Hectospec Spectroscopy}

The Hectospec instrument \citep{hectospec} mounted on the MMT 6.5m
telescope is ideal  for studying cluster infall regions
at moderate redshift.  Hectospec is a multiobject fiber-fed
spectrograph with 300 fibers deployable over a circular field-of-view
with a diameter of 1$^\circ$.  One Hectospec pointing extends to 
a radius of 2.3 (5.6) $\Mpc$ at $z$=0.1 (0.3), so a single Hectospec 
pointing covers the entire virial region and extends well into the infall region.

Our observing strategy for HeCS was to identify the red sequence of
cluster galaxies for each system and target primarily galaxies within 
0.3 magnitudes of the red sequence.  As shown in $\S \ref{redsequence}$, 
the color selection is significantly broader than the actual red sequence
of confirmed cluster members.  Thus, the target selection includes primarily, 
but not exclusively, red-sequence galaxies. 
We use SDSS DR6 spectroscopic data to identify galaxies with
existing redshifts and remove them from the target catalog.  The
galaxy targets have $r$ magnitudes of $r$=16-21.  The Hectospec fiber
assignment software xfitfibs\footnote{\url{https://www.cfa.harvard.edu/~john/xfitfibs/}} allows the user to rank
targets with priorities.  We ranked galaxies primarily by their
proximity to the cluster center and secondarily by apparent magnitude.
This procedure yields a largely complete magnitude-limited sample of
brighter galaxies supplemented by a more sparsely sampled selection of
galaxies up to 1 mag fainter than the bright limit.  The Appendix describes 
the target selection procedure in more detail. 

The galaxy targets are relatively bright for Hectospec spectroscopy:
we can obtain high-quality spectra with 3x20-minute exposures even under
suboptimal observing conditions (e.g., poor seeing, thin clouds).  Because Hectospec is a
queue-scheduled instrument, this flexibility allows the HeCS fields to
be observed under many observing conditions and improve the overall
efficiency of the Hectospec queue.  Observations were conducted 
primarily between 2007 June and 2009 February with a total of 10 
nights of queue time. We also observed one additional Hectospec field in A2261 
in 2011 May and seven additional fields in RXJ2129 in 2011 September.
We  observed additional fields in these clusters because
they are part of the CLASH sample \citep{postman12,coe12}.  The 
supplementary Hectospec fields targeted galaxies of all colors, enabling 
us to test the potential impact of our red-sequence selection on the 
estimates of dynamical parameters ($\S\ref{bluegalaxies}$).

Because the number of galaxies per cluster varies significantly, we
adjusted the limiting magnitudes primarily to obtain large, nearly
complete, samples to the faintest magnitude possible.  Thus, the
limiting absolute magnitude varies significantly from cluster to cluster.

After processing and reducing the spectra, we used the IRAF package 
{\it rvsao} \citep{km98} to cross-correlate the spectra with galaxy templates 
assembled from previous Hectospec observations.  During the pipeline 
processing, spectral fits are assigned a quality flag of ``Q" for high-quality 
redshifts, ``?" for marginal cases, and ``X" for poor fits.  Repeat observations 
of several targets with ``?" flags show that these redshifts are generally 
reliable.  We visually inspected all spectra to include ``?" and ``X" spectra 
that have redshifts secured by multiple lines (usually four or more).  Repeat 
observations indicate that the redshift uncertainties from {\it rvsao} are 
reasonable:  \citet{geller12a} use 1468 unique pairs of repeat observations 
to estimate a mean internal error of 56~$\kms$ for absorption-line objects and 
21~$\kms$ for emission-line objects \citep[see also][]{hectospec}.

Here we include new redshifts for 57 of the HeCS clusters.  For the 
remaining cluster, RXJ1720+26, \citet{owers11} obtained redshifts from 
MMT/Hectospec and Keck/DEIMOS for a separate investigation of 
cool-core clusters.  The data for RXJ1720+26 do not extend to the full 
field of Hectospec, but the spectra do extend to fainter apparent 
magnitudes than HeCS.   

We observed clusters at $z$$>$0.15 
with two configurations to obtain larger samples and to 
mitigate issues with fiber collisions.  We observed clusters at $z$=0.10-0.15 with one Hectospec configuration.
Table \ref{hecsredshifts} lists the \nczhecsnew unique redshifts obtained for HeCS. 
Columns [1]-[2] provide celestial coordinates, Column [3] is the heliocentric redshift $cz$, 
Column [4] is the uncertainty in $cz$, Column [5] is the cross-correlation score R, 
Column [6] is the quality flag, and Column [7] is a membership flag indicating the 
number of clusters for which this galaxy is classified as a member.  
In our analysis, we also use several thousand redshifts from SDSS 
(mostly foreground galaxies) and
132 redshifts around A2219 from \citet{boschin04}.  For convenience, 
Table \ref{hecsmemredshifts} provides these redshifts for the 2,621 
galaxies classified as members.  In total, we identify \nczmem galaxies
as members of a HeCS cluster.  There are 334 galaxies that are members of 
two HeCS clusters (i.e., the infall regions of a few clusters overlap slightly). 

\begin{table*}[th] \footnotesize
\begin{center}
\caption{\label{hecsredshifts} \sc HeCS Redshifts and Membership Classification}
\begin{tabular}{ccrrccc}
\tableline
\tableline
\tablewidth{0pt}
\multicolumn{2}{c}{Coordinates} & $cz_\odot$ & $\sigma_{cz}$  & R & Flag & Member \\ 
 RA (J2000) & DEC (J2000) &  km/s &km/s & &  & \\ 
\tableline
     1:50:49.556 &     1:02:12.196 & 69153 & 35 & 10.56 & Q & 0 \\ 
     1:50:50.707 &     1:04:06.564 & 113666 & 33 & 13.70 & Q & 0 \\ 
     1:50:52.011 &     0:52:56.928 & 122005 & 36 & 12.00 & Q & 0 \\ 
     1:50:52.308 &     1:05:49.391 & 188459 & 23 & 13.49 & Q & 0 \\ 
     1:50:52.987 &     0:58:50.880 & 106822 & 48 & 08.77 & Q & 0 \\ 
     1:50:53.390 &     0:53:45.420 & 68681 & 38 & 11.53 & Q & 1 \\ 
\tableline
\end{tabular}
\tablecomments{Table \ref{hecsredshifts} is published in its entirety in the electronic edition of the Journal. A portion is shown here for guidance regarding its form and content.}
\end{center}
\end{table*}

\begin{table*}[th] \footnotesize
\begin{center}
\caption{\label{hecsmemredshifts} \sc HeCS Members from Literature Redshifts }
\begin{tabular}{ccrrcc}
\tableline
\tableline
\tablewidth{0pt}
\multicolumn{2}{c}{Coordinates} & $cz_\odot$ & $\sigma_{cz}$  & Reference & Member \\ 
 RA (J2000) & DEC (J2000) &  km/s &km/s & \\ 
\tableline
 1:52:17.93 & 01:04:57.65 & 67795 & 51 & 1 & 1 \\ 
 1:52:50.47 & 01:12:45.11 & 68206 & 48 & 1 & 1 \\ 
 7:56:35.16 & 54:04:28.60 & 30795 & 42 & 1 & 1 \\ 
 7:58:18.82 & 53:57:37.69 & 30630 & 42 & 1 & 1 \\ 
 7:58:44.16 & 54:00:46.55 & 31295 & 39 & 1 & 1 \\ 
\tableline
\end{tabular}
\tablecomments{Table \ref{hecsmemredshifts} is published in its entirety in the electronic edition of the Journal. A portion is shown here for guidance regarding its form and content.}
\tablecomments{References: [1] SDSS, [2] Boschin et al.~2004.}
\end{center}
\end{table*}

\section{Results}

\subsection{Ubiquity of Infall Patterns around Clusters}

We first search for well-defined infall patterns around X-ray
clusters.  Analogous to CIRS, we plot the rest frame line-of-sight velocity relative to the
cluster center as a function of projected radius in Figure \ref{allhecs1}. All 58 HeCS systems
have ``clean" infall patterns; that is, there is little ambiguity in
the location of the caustics or limits of the pattern in redshift.
All clusters contain a large number of
members at the cluster redshift extending out to several Mpc from the
cluster center.  

\begin{figure}
\plotone{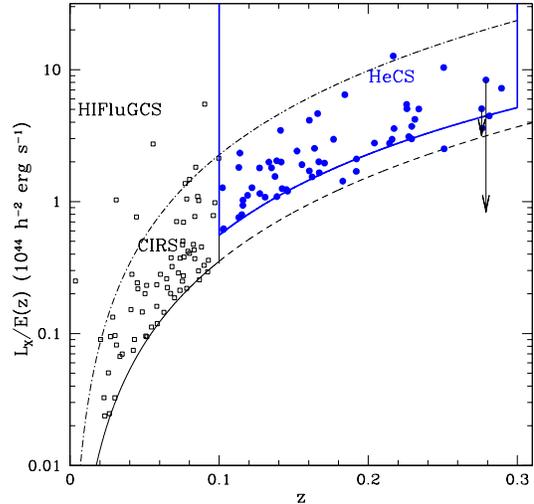}
\caption{\label{hecslxz} Redshift versus X-ray luminosity (0.1-2.4 keV) 
for X-ray clusters from CIRS (small open squares) and HeCS clusters
(filled blue circles) contained in the SDSS DR6 imaging survey
region.    
The blue solid lines show the selection of the HeCS sample: redshifts
0.1$<$$z$$<$0.3 and a flux limit of
$f_X>$5$\times$10$^{-12}$erg s$^{-1}$.  Four HeCS clusters are 
from a subsample of clusters with RA$>$17$^h$ and flux 
$f_X>$3$\times$10$^{-12}$erg s$^{-1}$ (the same flux limit as CIRS).
The lower solid line
shows the flux and redshift limits of the CIRS cluster sample.  The
dash-dotted line shows the flux limit 
(2 $\times$10$^{-11}$erg s$^{-1}$) of the HiFluGCS sample. }
\end{figure}

\begin{figure*}
\plotone{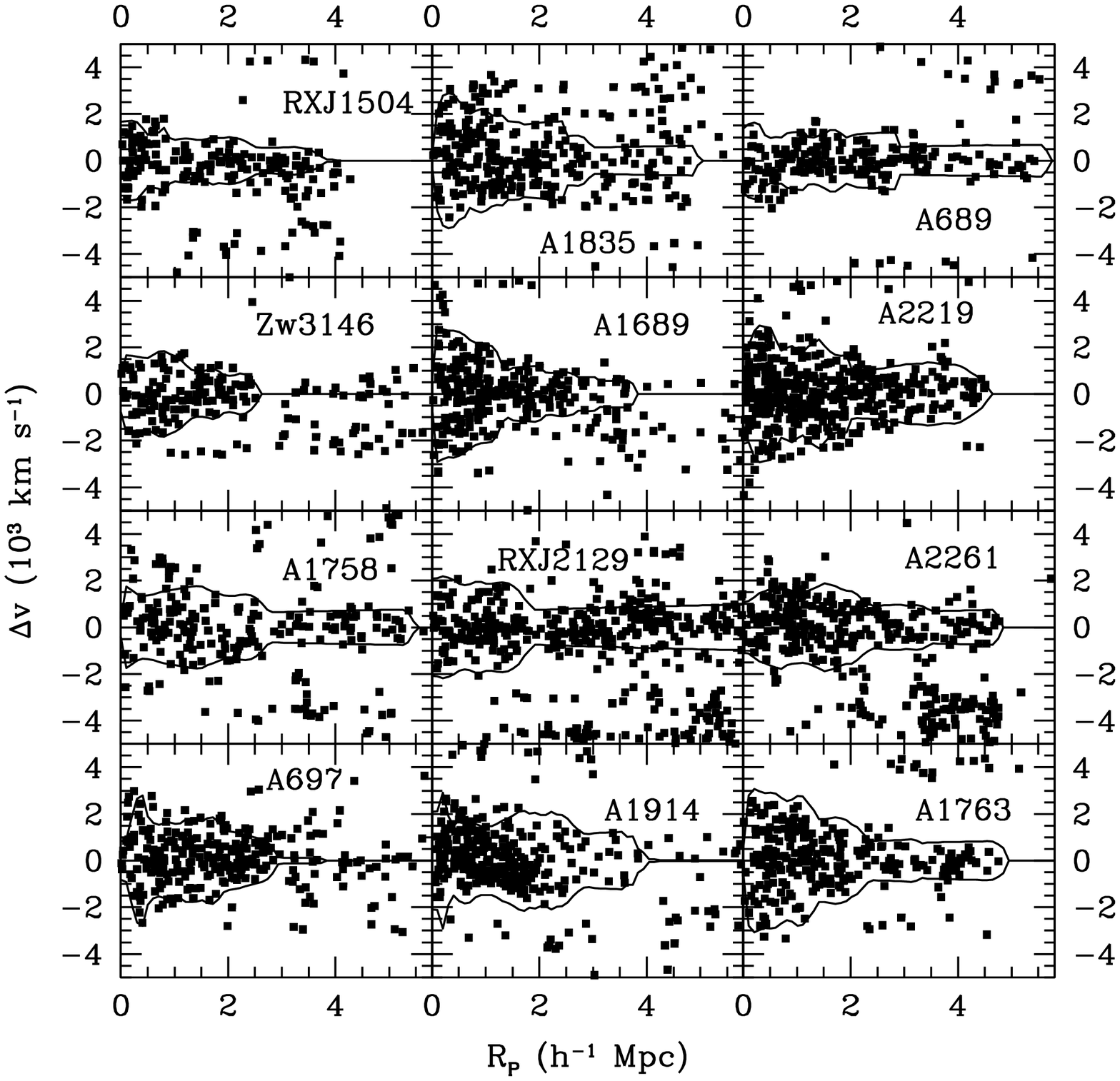}
\caption{\label{allhecs1} Redshift (rest-frame clustrocentric velocity) versus projected radius for galaxies around
HeCS clusters.  The caustic
pattern is evident as the trumpet-shaped regions with high density.
The solid lines indicate our estimate of the location of the caustics
in each cluster.  Clusters are ordered left-to-right and top-to-bottom
by decreasing X-ray luminosity. }
\end{figure*}

Figure \ref{hecslxz} shows the X-ray luminosities of the HeCS clusters
compared to HIFluGCS \citep{hiflugcs} and CIRS.  Due to the 
larger survey volume of HeCS, 
it includes many more systems with
$L_X$$\gtrsim$10$^{44} h^{-2}$erg s$^{-1}$ than the CIRS sample.
There is a substantial overlap at
intermediate $L_X$, especially at $z$=0.10-0.15.
Table \ref{sample} lists the clusters in the HeCS sample, their X-ray
positions and luminosities, their central redshifts and rest-frame velocity
dispersions.  Velocity dispersions are measured for galaxies within 
$r_{200}$ as determined from the caustic mass profiles.
The radius $r_\Delta$ is the radius within which the enclosed average
mass density is $\Delta\rho _c$ (where $\rho _c$ is the critical
density) by computing the enclosed density profile [$\rho (<r) = 3
M(<r)/4\pi r^3$]; $r_{200}$ is the radius which satisfies $\rho
(<r_{200})=200\rho_c$.

Figure \ref{allhecs1} shows the infall patterns and caustics for the 
HeCS clusters.  The contrast in phase space density between cluster 
members and foreground/background galaxies is striking. 

The clusters are ordered by decreasing X-ray luminosity.  
Velocity dispersions of clusters can be inferred visually from the spread 
in velocities at small radius.  One immediate conclusion is that velocity 
dispersion and X-ray luminosity are not perfectly correlated: clusters such 
as RXJ1504 and A689 have large X-ray luminosities, but their velocity 
dispersions are smaller than many clusters with comparable X-ray 
luminosity.  Cooling cores of varying X-ray luminosity contribute to scatter in 
the $L_X-\sigma_p$ relation \citep[e.g.,][]{zhang11a}.

\subsection{Caustics and Mass Profiles}

We calculate the shapes of the caustics with
the technique described in D99 using a smoothing parameter of $q=25$.
The smoothing parameter $q$ is the scaling between the velocity
smoothing and the radial smoothing in the adaptive kernel estimate of
the underlying phase space distribution.  Previous investigations show
that the mass profiles are insensitive to changes of a factor of 2 in
the smoothing parameter \citep{gdk99,rines2000,rines02}.  

The technique of D99 uses the redshifts and coordinates of the galaxies
to determine a hierarchical center based on a binary tree analysis. 
Analysis of 3000 simulated clusters indicates that the binary tree 
analysis recovers the input cluster center to within $\sim$300$\kpc$
for 95\% of simulated clusters \citep{serra11}.
For the HeCS clusters, the hierarchical centers are located within 
300$\kpc$ of the X-ray 
coordinates for all but four clusters (7\% of the sample, consistent 
with the 5\% of simulated clusters with offsets this large).
We discuss these four clusters and some other
individual cases in $\S$\ref{individual}.  For one cluster (A2261), 
we use the slightly different algorithm for cutting the binary tree 
described by \citet{serra11} to determine the center. 

Figure \ref{allhecs1} shows the phase-space diagrams (rest-frame 
velocity versus projected radius) and the caustics.   The D99 algorithm we 
use to identify the caustics generally
agrees with the lines one would draw based on a
visual impression. This consistency suggests that systematic
uncertainties in the caustic technique are dominated
by projection effects rather than the details of the
algorithm \citep[see][]{serra11}.

\begin{figure*}
\plotone{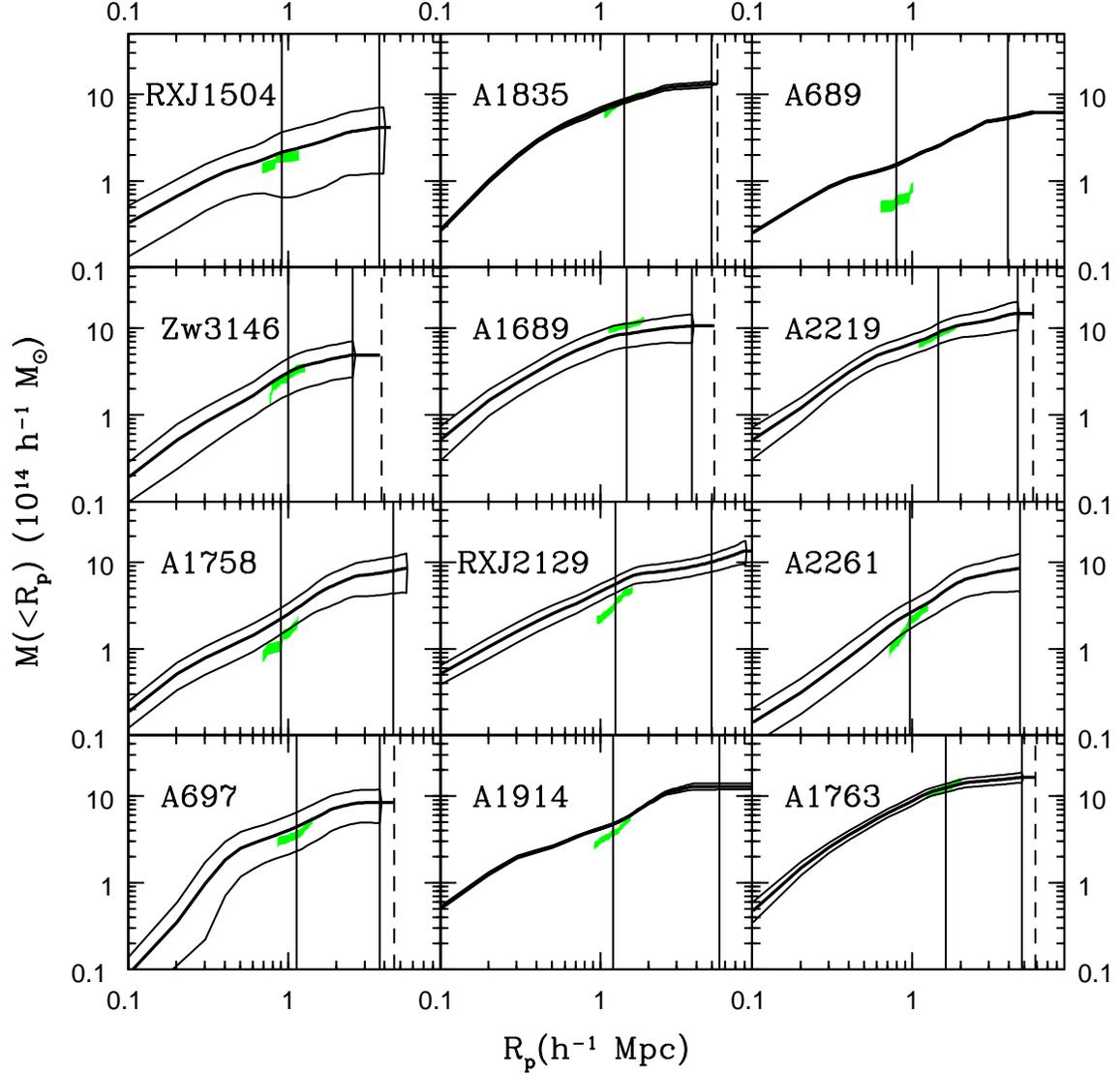}
\caption{\label{allhecsm1} Caustic mass profiles for the HeCS  clusters.
The thick solid lines show the caustic mass profiles and the thin
lines show the 1$\sigma$ uncertainties in the mass profiles. The inner vertical 
solid line in each panel shows the radius $r_{200}$.  The next vertical 
line shows the smaller of  $r_{5.6}$ (the limit of bound structure) and 
$r_{max}$ (the maximum radius where the caustics are detected). 
For clusters with $r_{max}<r_{5.6}$, a dashed vertical line shows a 
lower limit on $r_{5.6}$ assuming no mass is present outside $r_{max}$.
Green shaded regions indicate the virial mass profile in the range
(0.75-1.3)$r_{200}$ (approximately from $r_{500}$ to $r_{100}$). 
}
\end{figure*}

Figure \ref{allhecsm1} shows the associated caustic mass profiles.
In redshift space, a cluster of galaxies appears as trumpet-shaped 
pattern \citep[][]{rg89,1993ApJ...418..544V}.  DG and D99 demonstrated that for clusters 
forming hierarchically,  the boundaries of this sharply defined 
pattern (termed caustics) in redshift space (phase space) can be 
identified with the escape velocity from the cluster. This identification 
provides a route to estimation of the cluster mass profile assuming 
spherical symmetry. This mass estimation method is called the 
caustic method. 

The amplitude of the caustics $A(r)$ is half the distance between 
the boundaries of the cluster in redshift space. With the assumption 
of spherical symmetry the gravitational potential $\phi(r)$ and the
caustic amplitude $A(r)$ are related by

\beqn
 A^2(r) = -2\phi(r){{1 - \beta(r)} \over {3 -2\beta(r)}}
\eeqn
 
\noindent where $\beta(r)$ is the anisotropy parameter, $\beta(r) =  1 - \sigma_\theta^2(r)/[2\sigma_r^2(r)]$ where $\sigma_\theta$ and $\sigma_r$ are, respectively, the tangential and radial velocity dispersions.

DG show that the mass of a spherical shell within the infall region is the integral of the square of the caustic amplitude $A(r)$:

\beqn
\label{causticstomass}
GM(<r) - GM(<r_0) =  \mathcal{F}_\beta \int_{r_0}^{r} A^2(x)dx
\eeqn

\noindent where $ \mathcal{F}_\beta \simeq 0.5$ is a filling factor with a value estimated 
from numerical simulations. We approximate $ \mathcal{F}_\beta$ as a constant; variations 
in $ \mathcal{F}_\beta$ with radius lead to some systematic uncertainty in the mass profile 
we derive from the caustic technique. In particular, the caustic mass profile 
assuming constant $ \mathcal{F}_\beta$ somewhat overestimates the true mass profile 
within $\sim0.5\Mpc$ in simulated clusters \citep{serra11}.  We include these
issues in our assessment of the intrinsic uncertainties and biases in the 
technique \citep{serra11}.  

Some investigators have experimented with a 
modified caustic technique utilizing a radially-dependent $ \mathcal{F}_\beta(r)$ 
tailored to match simulated clusters \citep[e.g.,][]{bg03,lemze09}.  Because one goal of 
measuring mass profiles with the caustics is to test the predicted mass profiles 
from simulations, our approach is to assume a constant $ \mathcal{F}_\beta$ rather than 
impose a functional form for $ \mathcal{F}_\beta(r)$.

Note that the caustics extend to different
radii for different clusters. These differences result in part from the
varying physical size subtended by the Hectospec field at different redshifts.

D99 and \citet{serra11} show that the appearance of the caustics depends
strongly on the line of sight; projection effects can therefore
account for most of the differences in profile shape in Figure
\ref{allhecsm1} without invoking non-homology among
clusters.  

We use the caustics to determine cluster membership.  Here,
the term ``cluster member'' refers to galaxies both in the virial
region and in the infall region.  Figure
\ref{allhecs1} shows that the caustics effectively
separate cluster members from background and foreground galaxies.
Some interlopers unavoidably lie within the caustics \citep[e.g.,][]{serra11}. 
\citet{serra12} 
applied the 
caustic technique to 3000 clusters extracted from N-body simulations. 
The technique identifies 95$\pm$3\% of the true cluster members within 
$3r_{200}$. Only 2\% of the galaxies inside the caustics and
projected within $r_{200}$ are interlopers; the fraction
of interlopers reaches 8\% at $3r_{200}$.

\subsection{Comparison to Virial Mass Estimates \label{virialsec}}

\citet{zwicky1933,zwicky1937} first used the virial theorem to estimate the
mass of the Coma cluster.  With some modifications, notably a
correction term for the surface pressure \citep{1986AJ.....92.1248T},
the virial theorem remains in wide use \citep[e.g.,][and references
therein]{girardi98}.  Jeans analysis incorporates the radial
dependence of the projected velocity dispersion \citep[e.g.,][and
references therein]{cye97,2000AJ....119.2038V,bg03} and obviates the
need for a surface term.  Estimates of the mass profiles are complicated 
by the degeneracy between the mass and the velocity anisotropy 
\citep[e.g.,][and references therein]{mamon12}.

Virial mass estimators rely on the assumption that galaxy orbits are
in equilibrium, an assumption that is certainly violated in the infall region
and probably also in the inner regions of clusters with significant substructure.
Nevertheless, we apply the virial mass estimator to the HeCS clusters to 
check our caustic mass estimates.  
We must define a radius of
virialization within which the galaxies are relaxed.  We use $r_{200}$
as defined from the caustic mass profile (Table \ref{radii}) and include only galaxies within the caustics.  We
thus assume that the caustics provide a good division between cluster
galaxies and interlopers (see Figure \ref{allhecs1}).

We calculate the virial mass according to
\begin{equation}
M_{vir} = \frac{3 \pi}{2} \frac{\sigma_p^2 R_{PV}}{G}
\end{equation}
where $R_{PV} =  2N(N-1)/\sum_{i,j>i}R_{ij}^{-1}$ is the projected
virial radius and $\sigma_p^2 = \sum_i [(v_i-\bar v)/(1+\bar{z})]^2/(N-1)$. 
If the system does not lie entirely within $r_{200}$, a surface
pressure term 3PV should be added to the usual virial theorem so that
$2T + U = 3PV$. The virial mass is then an overestimate of the mass
within $r_{200}$ by the fractional amount 
\begin{equation}
\label{virialc}
C =  4\pi r_{200}^3 \frac{\rho (r_{200})}{\int_{0}^{r_{200}}4\pi r^2 \rho dr} \Bigg[{\frac{\sigma _{r} (r_{200})}{\sigma (<r_{200})}\Bigg]^2}
\end{equation}
where $\sigma_r (r_{200})$ is the radial velocity dispersion at
$r_{200}$ and $\sigma (<r_{200})$ is the enclosed total velocity
dispersion within $r_{200}$ \citep[e.g.,][]{girardi98}.  In the
limiting cases of circular, isotropic, and radial orbits, the maximum
value of the term involving the velocity dispersion is 0, 1/3, and 1
respectively. We estimate the uncertainties using the limiting fractional uncertainties $\pi^{-1} (2 \mbox{ln}
N)^{1/2}N^{-1/2}$.  These uncertainties do not include systematic
uncertainties due to membership determination.  Table \ref{radii} lists 
the virial and caustic mass estimates at the radius $r_{200}$ 
determined from the caustic mass profile.

\begin{figure}
\plotone{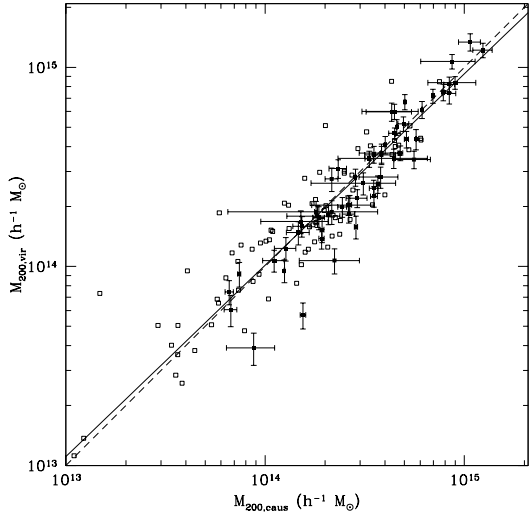} 
\caption{\label{hecsvc} Caustic masses at $r_{200}$ (determined from 
the caustic mass profile) compared to 
virial masses at the same radius.  Solid squares show HeCS clusters
and open squares show clusters from CIRS.  Errorbars show 1$\sigma$
uncertainties and the dashed line has slope unity. The solid line is the 
bisector of the ordinary least-squares fits. 
}
\end{figure}

Figure \ref{hecsvc}  compares the virial and caustic mass estimates at
$r_{200}$.  The mean ratios of these estimates are
$M_c/M_v = 1.12 \pm 0.04$. 
The caustic
mass estimates are slightly larger than virial mass estimates.  Including 
a correction factor $C\approx 0.1-0.2 M_{vir}$, consistent with the best-fit
NFW profiles \citep[see also ][CIRS]{cye97,girardi98,kg2000,cairnsi}, 
would lead to a larger difference.  
For the CIRS clusters, 
this ratio is slightly below unity.  When we combine the CIRS and 
HeCS cluster samples (130 total), we obtain an average of $M_c/M_v=1.011^{+0.033}_{-0.031}$. 
These results indicate that the caustic mass profile and the virial theorem 
yields similar masses on the scale of the virial radius (approximately $r_{200}$).


Figure \ref{allhecsm1} compares the mass profiles
estimated from the caustics and the virial theorem. 
The virial mass profiles are simply the virial mass estimator
applied to all galaxies (inside the caustics) within that 
projected radius.  Because the virial theorem only applies in regions
where the galaxies are in equilibrium, we only display
the virial mass profiles in the range $(0.75-1.3)r_{200}$ 
(corresponding roughly to $r_{500}$ to $r_{100}$).
The virial and caustic mass profiles generally agree. 
That is, near the virial radius, the caustic mass profiles do not appear 
to consistently overestimate or underestimate the mass
relative to the virial mass profiles.  This result supports our use of
caustic mass profiles as a tracer of the total cluster mass profile. 

\subsection{Virial  Masses and  Ultimate Halo Masses}

The caustic mass profiles allow direct estimates of the virial and
turnaround radius in each cluster.  For the virial radius, we estimate
$r_{200}$.  In our adopted cosmology, a system should be
virialized inside the slightly larger radius $\sim$$r_{100} \approx
1.3 r_{200}$ \citep{ecf96}.  We use $r_{200}$ because it is more
commonly used in the literature and thus allows easier comparison of
results.  

\citet{nl02} show that particles within a radius enclosing an overdensity 
of $\delta_c$=17.6 (enclosed density 5.56$\rho_{crit}$ or $\frac{9\pi^2}{16}\rho_{crit}$) in the present epoch remain bound to the central 
halo in numerical simulations of the far-future evolution of large-scale structure
in a $\Lambda$CDM universe. 
This criterion is further supported by simulations by other investigators
\citep{busha03,busha05,dunner06}.  To be precise, \citet{dunner06} 
show that about 10\% of the particles within this radius eventually become 
unbound, but that more distant particles constituting 13\% of the mass 
within $\delta_c$=17.6 are ultimately accreted by the halo; thus, the mass 
within $\delta_c$=17.6 is 1.03 times smaller than the mass of the halo 
when the scale factor is $a$=100\footnote{In previous work, we used the slightly 
more generous definition of the turnaround radius $r_{turn}$ determined from equation (8) of
\citet{rg89} assuming $\Omega _m = 0.3$.  For this value of $\Omega
_m$, the enclosed density is 3.5$\rho _c$ at the turnaround radius, versus 5.56$\rho_c$ for $\delta_c$=17.6.}.
If the $w$ parameter in the equation of state of the dark energy
($P_\Lambda = w\rho_\Lambda$) satisfies $w\ge -1$, the dark energy has
little effect on the turnaround overdensity
\citep[][]{2002MNRAS.337.1417G}.  Varying $\Omega_m$ in the range
0.02--1 only changes the inferred value of $r_{turn}$ by $\pm$10\%; the
uncertainties in $r_{turn}$ from the uncertainties in the mass profile are
comparable or larger \citep[D99;][]{rines02}.  
\citet{busha05} quantify the ultimate mass of dark matter haloes
in their simulations by the ratio $M_{5.6}/M_{200}$, i.e., the ratio of the 
ultimate mass to the present value of $M_{200}$.  They find that the 
mass ratio follow a log-normal distribution with a peak at 2.2 and a 
dispersion of 0.38 (about 10\% of halos occupy a high-end tail due to 
haloes merging with larger halos).  

\begin{figure}
\plotone{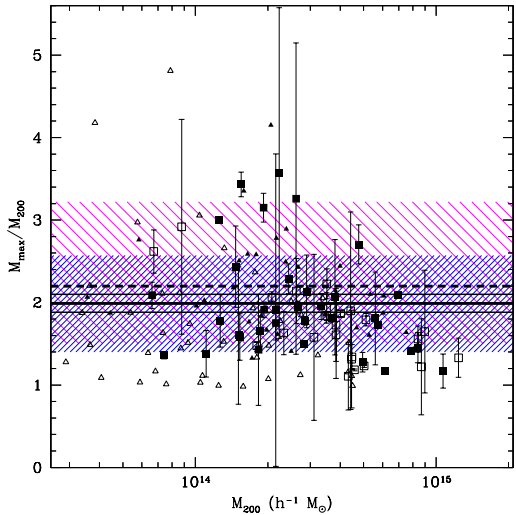}
\caption{\label{ultimatem} Ratio of mass $M_{max}$ within the maximum radius of the caustics (or $r_{5.6}$) to the mass $M_{200}$ within $r_{200}$.  Filled squares show clusters for which $r_{max}\geq r_{5.6}$ and open squares show clusters with $r_{max}<r_{5.6}$.  The thick solid line shows the mean value of $M_{5.6}/M_{200}$ for clusters with $r_{max}\geq r_{5.6}$. Thin solid lines show the uncertainty in this value.  Triangles show CIRS clusters. The dashed line at $M_{5.6}$=2.2$M_{200}$ is the ultimate mass of a halo in the far future (when the scale factor is $a$=100) compared to the present-day mass $M_{200}$ from the simulations of \citet{busha05}.  The distribution of $M(a=100)/M_{200}(a=1)$ in their simulations is well-described by a log-normal distribution with a dispersion shown by the magenta hatching (sloping down to the right).  The dispersion in $\ln(M_{5.6}/M_{200})$ for the HeCS clusters is shown by the dense blue hatching (sloping up to the right).}
\end{figure}

Table \ref{radii} lists $r_{200}$, $r_{5.6}$, and the masses $M_{200}$ and
$M_{5.6}$ enclosed within these radii.  For some clusters, the maximum
extent of the caustics $r_{max}$ is smaller than $r_{5.6}$.  For these
clusters, $M_{5.6}$ and $r_{5.6}$ are minimal values assuming that there is no
additional mass outside $r_{max}$.  The best estimate of the mass
contained in infall regions clearly comes from those clusters for
which $r_{max}\geq r_{5.6}$.  The average mass within 
$r_{5.6}$ for these clusters is 
1.99$\pm$0.11 times the virial mass
$M_{200}$, in remarkable agreement with the prediction of 
$2.2M_{200}$ \citep{busha05}.  Further, the dispersion in 
$\ln{M_{max}/M_{200}}$ is 0.30 (Figure \ref{ultimatem}), similar to the value of 
0.38 for simulated clusters \citep{busha05}.
The HeCS determination of $M_{5.6}/M_{200}$ demonstrates that 
clusters are still forming in the present epoch (this ratio is larger than unity).
Further, the measurement is consistent with the CIRS estimate of 2.19$\pm$0.18
(the CIRS estimate refers to $M_{3.5}$ rather than $M_{5.6}$; using 
$M_{5.6}$ for CIRS clusters would bring this value even closer to the HeCS estimate).

The remarkable agreement of the HeCS estimate of a cluster's 
ultimate halo mass with the prediction from simulations is a new 
test of $\Lambda$CDM structure formation theory.  

Because estimating the ultimate halo mass of a cluster (assuming a 
$\Lambda$CDM model) requires determining the mass profile 
to a radius of $r_{5.6}$,  the caustic technique is the only mass 
estimator that provides a direct probe of the ultimate halo mass.
Weak lensing can detect shear signals at these radii, but the 
contribution of line-of-sight structure is difficult to separate from the 
lensing shear of the galaxies bound to the cluster. 

\begin{figure}
\plotone{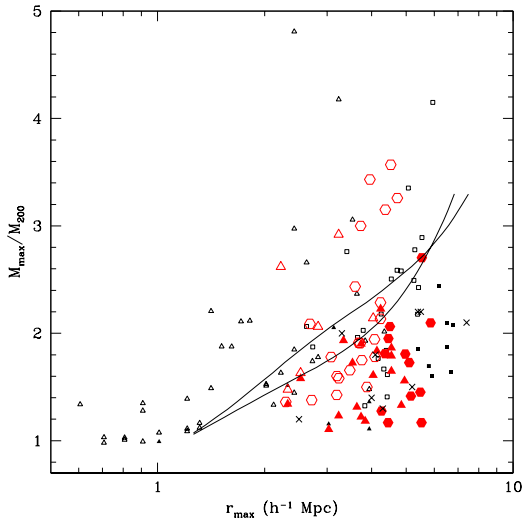}
\caption{\label{ctinker} Ratio of mass $M_{max}$ within the maximum radius of the caustics (or $r_{5.6}$) to $M_{200}$ versus the maximum radius $r_{max}$ of the caustics. Red hexagons are HeCS clusters with $M_{200}>3\times 10^{14}\msun$, red triangles are HeCS with $M_{200}<3\times 10^{14}\msun$.  Filled points have $r_{max}\geq r_{5.6}$.   Black points are CIRS clusters.
Filled squares show clusters for which $r_{max}\geq r_{5.6}$ and open squares show clusters with $r_{max}<r_{5.6}$.  See Figure \ref{ultimatem} for the typical uncertainties in $M_{max}/M_{200}$.  The lines show the mass profiles predicted by the simulations of \citet{tinker05}.  }
\end{figure}

The agreement between the estimates
of $M_{5.6}/M_{200}$ and $M_{3.5}/M_{200}$ between the CIRS and HeCS 
samples suggests that the overall shapes of cluster mass profiles into
their infall regions are not strongly dependent on cluster mass or redshift.  

In contrast, our mass profiles are perhaps in tension with the analysis of 
\citet{tinker05}.  Figure \ref{ctinker} compares our estimates of 
$M_{max}/M_{200}$ as a function of $r_{max}$ to the simulations of 
\citet{tinker05}.  Similar to CAIRNS and CIRS, the HeCS clusters tend to lie 
below the simulations, suggesting that either $M_{200}$ is overestimated
or that $M_{max}$ is often underestimated.  However, the $\sim$50\% 
offset suggested by \citet{tinker05} is difficult to reconcile with the 
good agreement between observed and simulated values of 
$M_{5.6}/M_{200}$ discussed above (see further discussion in $\S$ 4).

\begin{table*}[th] \footnotesize
\begin{center}
\caption{\label{radii} \sc HeCS  Characteristic Radii and Masses}
\begin{tabular}{lccrrrrrr}
\tableline
\tableline
\tablewidth{0pt}
Cluster & $r_{500}$ & $r_{200}$ & $r_{5.6}$ & $r_{max}$ & $M_{200}$ & $M_{vir}$ & $M_{5.6}$ & $M_{max}/M_{200}$ \\ 
 & $\Mpcx$ & $\Mpcx$ & $\Mpcx$ & $\Mpcx$ & $10^{14} M_\odot$ & $10^{14} M_\odot$ & $10^{14} M_\odot$ &  \\ 
\tableline
       A267 & 0.80 & 1.19 & 4.26 & 4.85 & 4.95$\pm$ 0.31 & 5.20$\pm$ 0.44 & 6.32$\pm$0.44 & 1.28$\pm$0.12   \\ 
     Zw1478 & 0.41 & 0.64 & 2.68 & 3.64 & 0.66$\pm$ 0.03 & 0.74$\pm$ 0.10 & 1.38$\pm$0.08 & 2.09$\pm$0.16   \\ 
       A646 & 0.67 & 0.98 & 4.24 & 10.00 & 2.44$\pm$ 0.12 & 1.99$\pm$ 0.23 & 5.58$\pm$0.34 & 2.29$\pm$0.24   \\ 
       A655 & 0.74 & 1.08 & 4.46 & 8.89 & 3.33$\pm$ 0.16 & 3.47$\pm$ 0.32 & 6.50$\pm$0.42 & 1.95$\pm$0.26   \\ 
       A667 & 0.73 & 1.02 & 4.09 & 7.68 & 2.86$\pm$ 0.05 & 1.58$\pm$ 0.21 & 5.09$\pm$0.12 & 1.78$\pm$0.06   \\ 
       A689 & 0.55 & 0.80 & 3.95 & 5.66 & 1.55$\pm$ 0.05 & 0.57$\pm$ 0.08 & 5.32$\pm$0.16 & 3.43$\pm$0.21   \\ 
       A697 & 0.76 & 1.13 & 4.60 & 3.74 & 4.42$\pm$ 2.10 & 3.49$\pm$ 0.39 & 8.40$\pm$3.48 & 1.90$\pm$1.20   \\ 
       A750 & 0.61 & 0.99 & 4.20 & 4.04 & 2.63$\pm$ 0.19 & 1.83$\pm$ 0.18 & 5.64$\pm$0.95 & 2.14$\pm$0.39   \\ 
     MS0906 & 0.41 & 0.81 & 3.59 & 3.94 & 1.47$\pm$ 0.19 & 1.48$\pm$ 0.20 & 3.58$\pm$0.55 & 2.44$\pm$0.52   \\ 
       A773 & 0.99 & 1.40 & 5.16 & 5.35 & 7.84$\pm$ 0.10 & 7.50$\pm$ 0.72 & 11.10$\pm$0.30 & 1.42$\pm$0.04   \\ 
       A795 & 0.70 & 1.10 & 4.44 & 4.14 & 3.52$\pm$ 0.07 & 3.65$\pm$ 0.35 & 6.47$\pm$0.15 & 1.84$\pm$0.06   \\ 
     Zw2701 & 0.57 & 0.86 & 3.19 & 3.23 & 1.83$\pm$ 0.54 & 1.79$\pm$ 0.22 & 2.61$\pm$0.96 & 1.43$\pm$0.67   \\ 
       A963 & 0.74 & 1.12 & 4.55 & 4.54 & 4.01$\pm$ 0.05 & 4.08$\pm$ 0.41 & 7.49$\pm$0.11 & 1.87$\pm$0.04   \\ 
       A980 & 0.80 & 1.18 & 4.27 & 3.64 & 4.46$\pm$ 1.40 & 6.00$\pm$ 0.53 & 5.86$\pm$1.89 & 1.31$\pm$0.59   \\ 
     Zw3146 & 0.60 & 1.00 & 3.83 & 2.52 & 3.11$\pm$ 1.41 & 2.62$\pm$ 0.33 & 4.91$\pm$2.20 & 1.58$\pm$1.01   \\ 
       A990 & 0.55 & 0.83 & 3.19 & 3.64 & 1.51$\pm$ 0.56 & 1.68$\pm$ 0.22 & 2.42$\pm$0.89 & 1.60$\pm$0.84   \\ 
     Zw3179 & 0.43 & 0.63 & 2.87 & 2.22 & 0.67$\pm$ 0.05 & 0.61$\pm$ 0.11 & 1.77$\pm$0.12 & 2.62$\pm$0.26   \\ 
      A1033 & 0.66 & 1.03 & 3.88 & 10.00 & 2.84$\pm$ 0.03 & 2.80$\pm$ 0.28 & 4.26$\pm$0.06 & 1.50$\pm$0.03   \\ 
      A1068 & 1.01 & 1.47 & 5.48 & 10.00 & 8.40$\pm$ 0.66 & 7.40$\pm$ 0.85 & 12.20$\pm$1.20 & 1.45$\pm$0.19   \\ 
      A1132 & 0.76 & 1.10 & 4.73 & 4.24 & 3.52$\pm$ 0.14 & 2.25$\pm$ 0.27 & 7.84$\pm$0.56 & 2.23$\pm$0.18   \\ 
      A1201 & 0.60 & 0.99 & 4.08 & 4.65 & 2.66$\pm$ 0.06 & 2.04$\pm$ 0.23 & 5.17$\pm$0.15 & 1.94$\pm$0.08   \\ 
      A1204 & 0.49 & 0.74 & 2.71 & 3.94 & 1.11$\pm$ 0.14 & 1.07$\pm$ 0.13 & 1.53$\pm$0.25 & 1.38$\pm$0.32   \\ 
      A1235 & 0.57 & 0.84 & 3.23 & 7.88 & 1.53$\pm$ 0.15 & 1.59$\pm$ 0.19 & 2.42$\pm$0.36 & 1.58$\pm$0.28   \\ 
      A1246 & 0.86 & 1.22 & 4.89 & 4.54 & 5.12$\pm$ 0.12 & 4.35$\pm$ 0.41 & 9.18$\pm$0.32 & 1.79$\pm$0.08   \\ 
      A1302 & 0.59 & 0.93 & 3.90 & 2.83 & 2.08$\pm$ 0.03 & 1.82$\pm$ 0.20 & 4.30$\pm$0.12 & 2.06$\pm$0.07   \\ 
      A1361 & 0.55 & 0.78 & 3.73 & 10.00 & 1.25$\pm$ 0.00 & 0.95$\pm$ 0.12 & 3.75$\pm$0.02 & 3.00$\pm$0.03   \\ 
      A1366 & 0.59 & 0.90 & 4.37 & 10.00 & 1.92$\pm$ 0.06 & 1.38$\pm$ 0.16 & 6.05$\pm$0.27 & 3.15$\pm$0.18   \\ 
      A1413 & 0.88 & 1.29 & 5.10 & 10.00 & 5.72$\pm$ 0.02 & 4.36$\pm$ 0.51 & 9.88$\pm$0.04 & 1.73$\pm$0.01   \\ 
      A1423 & 0.73 & 1.09 & 4.36 & 4.75 & 3.68$\pm$ 0.06 & 2.59$\pm$ 0.26 & 6.66$\pm$0.13 & 1.81$\pm$0.05   \\ 
      A1437 & 1.08 & 1.59 & 5.53 & 10.00 & 10.70$\pm$ 1.35 & 13.40$\pm$ 1.33 & 12.50$\pm$1.55 & 1.17$\pm$0.21   \\ 
      A1553 & 0.81 & 1.19 & 4.14 & 3.84 & 4.58$\pm$ 0.03 & 5.01$\pm$ 0.45 & 5.42$\pm$0.04 & 1.18$\pm$0.01   \\ 
      A1682 & 0.88 & 1.28 & 4.45 & 5.25 & 6.13$\pm$ 0.04 & 6.13$\pm$ 0.60 & 7.16$\pm$0.05 & 1.17$\pm$0.01   \\ 
      A1689 & 1.01 & 1.46 & 5.15 & 3.74 & 8.68$\pm$ 2.64 & 10.70$\pm$ 0.90 & 10.60$\pm$3.90 & 1.22$\pm$0.58   \\ 
      A1758 & 0.57 & 0.90 & 4.53 & 5.46 & 2.23$\pm$ 0.75 & 1.07$\pm$ 0.15 & 7.96$\pm$3.59 & 3.57$\pm$2.23   \\ 
      A1763 & 1.10 & 1.62 & 5.86 & 4.85 & 12.40$\pm$ 1.39 & 12.20$\pm$ 1.02 & 16.50$\pm$2.29 & 1.33$\pm$0.24   \\ 
      A1835 & 0.95 & 1.41 & 5.40 & 4.95 & 8.41$\pm$ 0.53 & 8.21$\pm$ 0.72 & 13.10$\pm$1.06 & 1.56$\pm$0.16   \\ 
      A1902 & 0.49 & 0.95 & 3.69 & 2.52 & 2.33$\pm$ 0.23 & 3.10$\pm$ 0.33 & 3.80$\pm$0.48 & 1.63$\pm$0.26   \\ 
      A1918 & 0.60 & 0.90 & 3.68 & 10.00 & 1.93$\pm$ 0.04 & 1.53$\pm$ 0.21 & 3.69$\pm$0.17 & 1.91$\pm$0.10   \\ 
      A1914 & 0.82 & 1.20 & 5.54 & 10.00 & 4.77$\pm$ 0.13 & 3.71$\pm$ 0.32 & 12.90$\pm$1.07 & 2.70$\pm$0.24   \\ 
      A1930 & 0.60 & 0.89 & 3.47 & 4.24 & 1.86$\pm$ 0.12 & 1.76$\pm$ 0.23 & 3.08$\pm$0.27 & 1.66$\pm$0.19   \\ 
      A1978 & 0.41 & 0.69 & 3.25 & 3.23 & 0.88$\pm$ 0.24 & 0.39$\pm$ 0.07 & 2.56$\pm$0.76 & 2.92$\pm$1.30   \\ 
      A2009 & 0.70 & 1.09 & 4.49 & 3.33 & 3.51$\pm$ 0.17 & 2.47$\pm$ 0.24 & 6.79$\pm$0.56 & 1.93$\pm$0.19   \\ 
    RXJ1504 & 0.61 & 0.91 & 3.71 & 3.94 & 2.16$\pm$ 1.51 & 1.88$\pm$ 0.26 & 4.12$\pm$2.91 & 1.91$\pm$1.90   \\ 
      A2034 & 0.81 & 1.25 & 4.41 & 3.23 & 5.03$\pm$ 0.05 & 6.71$\pm$ 0.58 & 6.20$\pm$0.07 & 1.23$\pm$0.02   \\ 
      A2050 & 0.76 & 1.18 & 4.04 & 3.03 & 4.32$\pm$ 1.11 & 5.98$\pm$ 0.62 & 4.78$\pm$1.28 & 1.11$\pm$0.41   \\ 
      A2055 & 0.61 & 0.94 & 3.75 & 10.00 & 2.16$\pm$ 0.16 & 2.75$\pm$ 0.37 & 3.78$\pm$0.34 & 1.75$\pm$0.27   \\ 
      A2069 & 0.85 & 1.39 & 5.86 & 10.00 & 6.96$\pm$ 0.08 & 7.18$\pm$ 0.58 & 14.60$\pm$0.20 & 2.10$\pm$0.04   \\ 
      A2111 & 0.69 & 1.00 & 4.24 & 4.75 & 2.90$\pm$ 0.35 & 2.20$\pm$ 0.23 & 6.19$\pm$1.04 & 2.13$\pm$0.47   \\ 
      A2187 & 0.52 & 0.77 & 3.08 & 4.24 & 1.27$\pm$ 0.16 & 1.23$\pm$ 0.17 & 2.26$\pm$0.30 & 1.78$\pm$0.32   \\ 
      A2219 & 0.97 & 1.46 & 5.67 & 4.54 & 8.98$\pm$ 2.42 & 8.37$\pm$ 0.62 & 14.80$\pm$5.36 & 1.65$\pm$0.74   \\ 
     Zw8197 & 0.57 & 0.89 & 3.33 & 2.32 & 1.80$\pm$ 0.03 & 1.88$\pm$ 0.24 & 2.66$\pm$0.06 & 1.48$\pm$0.04   \\ 
      A2259 & 0.77 & 1.12 & 4.44 & 3.54 & 3.84$\pm$ 0.68 & 3.63$\pm$ 0.38 & 6.63$\pm$1.23 & 1.73$\pm$0.44   \\ 
    RXJ1720 & 0.79 & 1.18 & 4.29 & 2.32 & 4.47$\pm$ 0.30 & 4.66$\pm$ 0.30 & 6.01$\pm$0.52 & 1.34$\pm$0.15   \\ 
      A2261 & 0.57 & 0.97 & 4.72 & 4.75 & 2.62$\pm$ 0.91 & 2.03$\pm$ 0.23 & 8.54$\pm$3.97 & 3.26$\pm$1.89   \\ 
    RXJ2129 & 0.80 & 1.24 & 4.97 & 8.08 & 5.59$\pm$ 1.16 & 3.44$\pm$ 0.34 & 10.10$\pm$2.34 & 1.81$\pm$0.89   \\ 
      A2396 & 0.74 & 1.11 & 4.29 & 4.04 & 3.84$\pm$ 0.87 & 3.71$\pm$ 0.41 & 6.19$\pm$1.47 & 1.61$\pm$0.53   \\ 
      A2631 & 0.70 & 1.07 & 4.50 & 5.25 & 3.80$\pm$ 0.84 & 2.81$\pm$ 0.34 & 7.84$\pm$2.01 & 2.06$\pm$0.77   \\ 
      A2645 & 0.43 & 0.63 & 2.29 & 3.43 & 0.74$\pm$ 0.01 & 0.92$\pm$ 0.13 & 1.01$\pm$0.03 & 1.36$\pm$0.04   \\ 
\tableline
\end{tabular}
\end{center}
\end{table*}

One striking result of this analysis is that the caustic pattern is
often visible beyond the radius $r_{5.6}$ that marks the maximum
radius of galaxies that are gravitationally bound to the cluster.  This result
suggests that clusters may have strong dynamic effects on surrounding
large-scale structure beyond the radius where galaxies will remain bound to the
cluster in the far future.  

\subsection{Cluster Scaling Relations} 

Scaling relations between simple cluster observables and masses provide
insight into the nature of cluster assembly and the properties of
various cluster components.  Establishing these relations for local
clusters is critical for future studies of clusters in the distant
universe with the goal of constraining dark energy
\citep{majumdar04,lin04}.  

We apply the prescription of \citet{danese} to determine the mean
redshift $cz_\odot$ and projected velocity dispersion $\sigma_p$ of
each cluster from all galaxies within the caustics.  We calculate
$\sigma_p$ using only the cluster members projected within $r_{200}$
estimated from the caustic mass profile.  Note that our estimates of
$r_{200}$ do not depend on $\sigma_p$.

\begin{figure}
\plotone{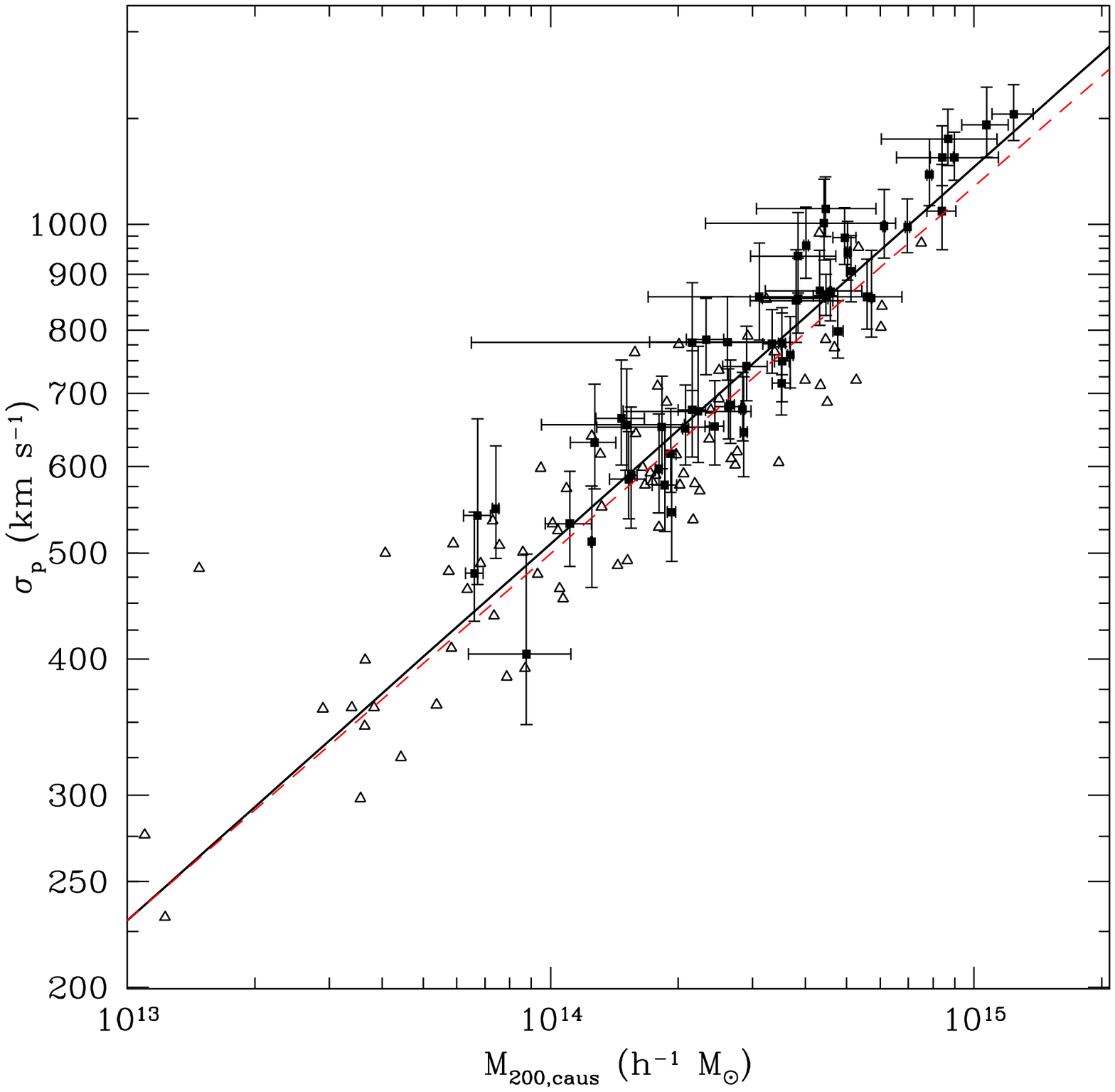}
\caption{\label{msigma} Caustic masses at $r_{200}$ compared to velocity 
dispersions within $r_{200}$.  Squares and triangles show HeCS and CIRS 
clusters respectively.  The solid line is the bisector of the ordinary least-squares 
fits.  The dashed line is the 
$\sigma_p-M_{200}$ relation of dark matter particles in cosmological
simulations by \citet{evrard07}.  }
\end{figure}

Figure \ref{msigma} shows the $M_{200} - \sigma _p$ relation.  The
tight relation indicates that the caustic masses are well correlated
with velocity dispersion estimates.  The good correlation is 
not surprising because both parameters depend on the galaxy velocity
distribution.  The best-fit slope is
$M_{200}\propto\sigma_p^{2.90\pm0.15}$ with the uncertainty estimated
from jackknife resampling.  
The dashed line in Figure \ref{msigma} shows the $M_{200}-\sigma_p$ 
relation found by \citet{evrard07} for dark matter particles in simulated 
dark matter haloes.  \citet{evrard07} find that this relation is insensitive to 
variations in cosmological parameters or numerical resolution 
(above $10^3$ tracer particles).  The excellent agreement between 
the observed CIRS and HeCS clusters and the virial scaling relations 
from simulated dark matter haloes (slope 2.98$\pm$0.02) suggests that the caustic technique 
yields accurate mass estimates.

\begin{figure}
\plotone{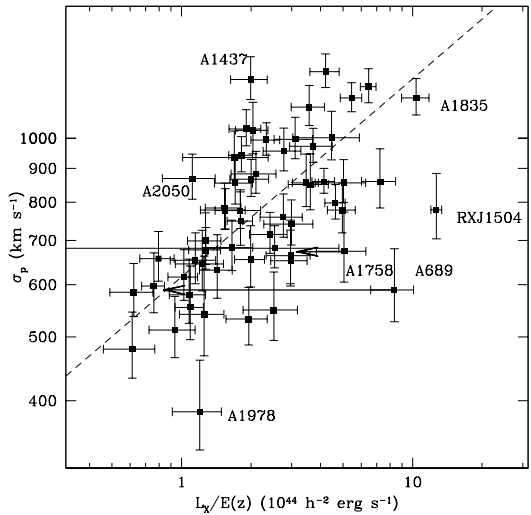}
\caption{\label{lxsigma} Rest-frame ROSAT X-ray luminosities compared to velocity 
dispersions within $r_{200}$.  Several outliers are labeled, and arrows indicate 
corrected luminosities for A689 and A1758.  The line shows the best-fit relation from
\citet{zhang11a} for $L_X$ in the {\em ROSAT} band and all emission within $r_{500}$ 
(i.e., before excluding emission from cooling cores).}
\end{figure}

Figure \ref{lxsigma} compares the HeCS velocity dispersions to rest-frame 
X-ray luminosities in the ROSAT band.   For reference, Figure \ref{lxsigma} shows the best-fit 
$L_X-\sigma_p$ relation from Table C.2 of \citet{zhang11a} for $L_X$ measured in the 
{\em ROSAT} band including all emission within $r_{500}$ (i.e., before excluding 
emission from cooling cores).  The HeCS clusters generally follow this relation, but there
are also several outliers. 
Figure \ref{lxsigma} labels several of these outliers. 
Two outliers, A689 and A1758, have problematic X-ray luminosities.  Figure 
\ref{lxsigma} shows that A689 lies close to the main locus of points when 
the X-ray luminosity is corrected for the central point source.  
\citet{zhang11a} show that the $L_X-\sigma_p$ relation has smaller scatter 
when cool cores are excised from the $L_X$ measurements.  Unfortunately,
most HeCS clusters lack the high-resolution X-ray imaging required for this analysis.
We defer a full
analysis of the $L_X-\sigma_p$ relation to future work.  

\subsection{Comments on Individual Clusters \label{individual}}

Clusters share many common features, but any large sample of clusters
contains some complex systems.  We comment on some of the most
exceptional cases here.  We present a comparison of caustic mass 
profiles and weak lensing mass profiles of several HeCS clusters
in \citet{geller12b}.

{\it A667.} --- The hierarchical center of A667 is located 307$h^{-1}$kpc 
N of the BCG.  The X-ray center is close to the position of the BCG.
The spatial distribution of cluster members shows significant substructure
to the North.  This substructure accounts for the offset in the centers.

{\it A689.} --- This cluster was classified in BCS as a compact X-ray source, 
suggesting possible contamination by a central point source \citep{bcs}.  A 
recent {\em Chandra} observation confirms that most of the X-ray luminosity 
in BCS is due to a central point source identified as a BL Lac \citep{giles11}.  
\citet{giles11} estimate that the cluster luminosity in the BCS catalog is 
overestimated by about a factor of 10.  With the revised luminosity, A689 
lies below the flux limit of our flux-limited sample (Figure \ref{hecslxz}).  The 
Hectospec redshifts confirm that the mass of A689 is significantly smaller 
than the masses of other clusters with similar (uncorrected) $L_X$.  Figure 
\ref{lxsigma} shows that A689 is not an outlier in the $L_X-\sigma_p$ 
diagram when using the corrected $L_X$.  This cluster highlights the 
importance of high-resolution X-ray observations in determining accurate 
X-ray luminosities.

\begin{figure}
\plotone{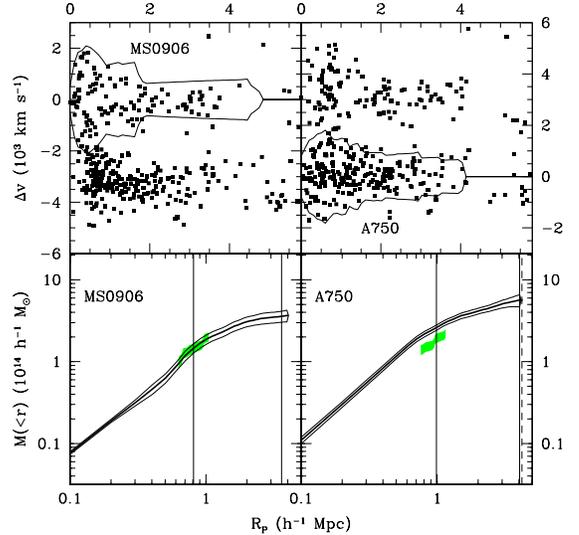}
\caption{\label{ms0906} {\it Top panels:} Redshift versus projected radius for 
MS0906 (left) and A750 (right).  The caustic pattern of A750 is much clearer 
when the plot is made with A750 at the center (cf. Figure \ref{allhecs1}). 
{\it Bottom panels:} Mass profiles of MS0906 (left) and A750 (right). }
\end{figure}

{\it MS0906/A750.} --- This pair of clusters is a curious system.  MS0906+11
is an X-ray cluster at $z$=0.1767 detected in the {\it Einstein} Medium 
Sensitivity Slew Survey \citep{henry92}.  A750 is a cluster at $z$=0.1640 
located only 5$\arcm$ (0.63$\Mpc$) away from the X-ray center of MS0906 
\citep[see Figure 3.39 of][]{maughan08}.  \citet{cnoc96} noted that MS0906 
``appears to be an indistinct binary in redshift space".   With the denser 
sampling of HeCS, the infall patterns of the two clusters are separable.  Figure 
\ref{ms0906} shows the caustics and mass profiles of these two clusters.  The two 
clusters are separated by 3250$\kms$ (rest-frame), suggesting that they 
are not gravitationally bound.  Note that the weak lensing map of 
\citet{okabe10} shows two distinct mass components centered 
approximately on MS0906 (component A750-C in their Figure 32) and A750 
(component A750-NW1 in their Figure 32); MS0906 has larger surface mass 
density and A750 has a larger luminosity density of red-sequence galaxies.
The X-ray luminosity of MS0906 is much larger than that of A750 \citep{maughan08}.
Our caustic mass profiles indicate that the two clusters have roughly equal 
mass.  This complex system 
indicates that the relation between cluster mass and different observables 
is complicated.  A more detailed analysis of this system is presented in 
\citet{geller12b}.

{\it A773.} --- A773 contains a radio halo \citep{giovannini99}, a feature 
commonly associated with major mergers.  \citet{barrena07} studied the 
dynamics of 100 cluster members and found evidence of complicated 
dynamics, including two velocity peaks separated by $\approx 2500\kms$.  
Each velocity peak contains one of the two bright galaxies lying in the cluster
center.  A third galaxy (SDSS J091758.60+515104.6) is intermediate in 
brightness to the two bright galaxies in the core; this third galaxy is in the 
lower-velocity peak and is relatively isolated.   The caustics contain 
both systems, although some galaxies in the high-velocity peak lie outside 
the caustics.
Because the caustic method does not require equilibrium, 
the caustic method may be a more robust mass estimator for A773 than 
virial analysis.  The hierarchical center has a redshift similar to the 
low-velocity peak, and the caustics at large radii are centered at the same redshift. 
Note that our estimate of the velocity dispersion (1110$^{+86}_{-70}\kms$) is 
smaller than the global velocity dispersion 1394$^{+84}_{-68}\kms$ of
\citet{barrena07}, although they find much smaller velocity dispersions of 
individual subclusters.  


{\it A1437.} --- This cluster was studied by \citet{pimbblet06}.  They measured 
a velocity dispersion of $1152^{+59}_{-51}\kms$, smaller than but consistent
with our estimate ($\sigma_p=1233^{+102}_{-81}\kms$).


{\it A1689.} --- This cluster is a famous strong-lensing cluster \citep[e.g.,][]{broadhurst05a}.  
\citet{lokas06b} studied the dynamics of A1689 from literature data and find significant 
line-of-sight substructure.  The dynamics have also been studied by \citet{lemze09} 
using more extensive VLT/VIMOS spectroscopy obtained by \citet{czoske04}.  
\citet{lemze09} show that the mass profile determined with the caustic technique
is consistent with mass profiles from gravitational lensing, X-ray data, and Jeans' analysis. 
\citet{haines10} use 1009 Hectospec redshifts (from an independent investigation)
to study the properties of star-forming galaxies detected by {\em Herschel}.  
Note that \citet{lemze09} claim to detect the ``edge" of A1689 by noting a 
sharp decrease in galaxy density at $R_p\approx 2.1\Mpc$.  Figure \ref{allhecs1}
shows several spectroscopically confirmed members beyond this radius,
and we detect caustics extending to $R_p=3.7\Mpc$. 


{\it A1758.} --- This system is a merger of multiple X-ray clusters at 
$z$=0.28.  Several authors separate the cluster into A1758N and 
A1758S \citep[e.g.,][]{bcs,david04}.  The combined flux of the two clusters 
\citep[][]{noras}
exceeds our flux limit; the individual fluxes lie below the flux limit.  
\citet{david04} showed that A1758N is itself a merger of two 7 keV clusters
and that A1758S  shows evidence of a recent merger. 
\citet{okabe08} used Subaru to analyze the lensing properties of 
A1758N and A1758S and found confirming evidence that both components 
are undergoing mergers.  \citet{ragozzine12} use a higher-resolution 
lensing analysis of A1758N and conclude that A1758N consists of 
two separate clusters A1758N:NW and A1758N:SE.   Because 
A1758 is one of the highest-redshift clusters in HeCS, it is not 
sampled very deeply.  A detailed dynamical analysis of this 
complex system would require additional data. The complex 
dynamics of the mergers in A1758 may contribute to its unusually 
high star formation rate \citep{haines09}.  \citet{boschin12} study the 
dynamics of A1758N:NW and A1758N:SE using TNG spectroscopy. 
They find a much larger velocity dispersion ($\sigma_p\sim 1300\kms$) 
than we do, but they note that many high-velocity-offset galaxies might 
be members of subclusters.  



{\it A1902.} --- The hierarchical center is coincident with a close 
grouping of galaxies 531$h^{-1}$kpc E of the BCG.  The close 
grouping is probably a merging event, leading to a very high 
binding energy that causes the hierarchical analysis to choose 
this grouping as the cluster center.  The X-ray center is located 
close to the BCG, and the spatial distribution of cluster members 
shows many more members East of the BCG than West of the 
BCG.  The BCG has some close companion galaxies 
that may have been missed by either SDSS photometry or by 
fiber collisions.  In either case, the BCG companion galaxies 
would be omitted from the redshift catalog used for the 
hierarchical analysis; including some of these companions 
could conceivably relocate the hierarchical center close to the BCG. 

{\it A1914.} --- Although the X-ray contours of A1914 are fairly 
regular, it contains a large radio halo and significant X-ray 
temperature anisotropy, two features consistent with a major 
merger \citep{govoni04}.   Similarly, a weak lensing map by  
\citet{okabe08}  shows significant substructure in both the lensing 
mass distribution and the luminosity distribution (which includes 
two bright galaxies close to but not coincident with the X-ray center).  The HeCS 
data (Figure \ref{allhecs1}) show a clear caustic pattern but 
with an unusual shape: the median redshift of cluster 
members increases from the center to $\sim 0.5\Mpc$, then 
decreases from this radius to $\sim 2.0\Mpc$.  Inspection of the
spatial distribution of the cluster members reveals that a 
large group is located $\approx 1.5\Mpc$ NW of the X-ray center.
The group is offset by $\approx -1000\kms$
from the redshift center of A1914 and has a velocity dispersion 
of $\sigma_p\sim350\kms$.  This group largely explains the 
unusual appearance of the infall pattern in Figure \ref{allhecs1}.  
We defer a 
full dynamical analysis of this complex system to future work. 

{\it A1930.} --- The hierarchical center of A1930 is located 385$h^{-1}$kpc 
from the X-ray center.  The hierarchical center is 3$\farcm$5 SSE 
of the BCG and 1$\arcm$ from the second-ranked galaxy.  The 
X-ray center is 2$\farcm$5 WSW of the BCG.  The BCG 
is located about 1$\farcm$5 NNE of the midpoint of the hierarchical center and 
the X-ray center. 

{\it RXJ1504-03.} ---  This cluster is near the edge of the DR6 photometric 
footprint.  Our target photometry is based on DR5, which did not contain 
the entire cluster.  Part of the Hectospec field was contained in DR5 imaging 
that did not yield photometry.  We used the SDSS Navigate tool to identify 
and include red and blue galaxies in our target list by inspecting the 
unprocessed SDSS imaging.  A small portion of the Hectospec field was 
covered by neither SDSS photometry nor Atlas imaging.  We added targets 
from POSSII plate scans  in this region.   

RXJ1504 is well known as a cooling-core cluster \citep{bohringer05}, and 
it is the most X-ray luminous cluster in HeCS.  Vikhlinin et al (2009) estimate 
a large mass for RXJ1504 based on {\em Chandra} observations and the 
$Y_X$ estimator.  However, Figures \ref{allhecs1} and \ref{lxsigma} show 
that its velocity dispersion is smaller than many clusters of comparable 
X-ray luminosity.  This result suggests that X-ray mass estimates of 
RXJ1504 (including $Y_X$) may be biased due to the cooling core.  
Recently, \citet{zhang12} studied the dynamics of galaxies in RXJ1504 
using VLT/VIMOS spectroscopy.  They identify 53 cluster members and 
compute a velocity dispersion of 
$\sigma_p$=(1132$\pm$95)$\kms$, a value significantly larger than ours ($\sigma_p$=779$^{+105}_{-75}\kms$, 
Table \ref{sample}).  Comparing their Figure 7 to our Figure \ref{allhecs1} 
indicates that their membership selection is more generous than our 
caustic selection; HeCS includes more redshifts (120 members versus 53) 
and covers a wider field-of-view than the spectra contained in \citet{zhang12}.  
In addition, \citet{zhang12} identify several blue galaxies as cluster 
members: if these galaxies have a larger velocity dispersion than the 
red galaxies we target, this difference could partly explain the difference 
in measured velocity dispersions.  However, analysis of blue galaxies in 
other HeCS clusters ($\S \ref{bluegalaxies}$) shows no significant color dependence of the velocity
distribution of members identified with large redshift samples and the 
caustic technique.

{\it A2055.} --- This cluster was studied by \citet{pimbblet06}.  They measured   
a velocity dispersion of $1046^{+80}_{-65}\kms$, significantly larger
than our estimate ($\sigma_p=676^{+90}_{-64}$).  Inspection of the
infall pattern shown in their Figure 5 reveals that they classify three 
galaxies with large velocity offsets as members; similar galaxies are 
classified as non-members by the caustic technique (Figure \ref{allhecs1}).


{\it A2219.} --- The kinematics of this well-known lensing 
cluster \citep{smail95} were studied 
by \citet{boschin04} using 132 redshifts within 5$\arcm$ of the BCG.  
They find significant evidence of substructure and possible merging 
activity. 
We include their redshifts in our caustic analysis 
and we removed these galaxies from our Hectospec target list.  A2219 
contains a radio halo \citep{giovannini99}.  Figure \ref{allhecs1} shows 
that there are a few galaxies projected in front of A2219; while we 
classify these galaxies as foreground, \citet{boschin04} classify 
them as members.  This membership difference probably accounts 
for most of the difference between their estimated velocity dispersion
($\sigma_p=1438^{+109}_{-86}\kms$) and our estimate (Table \ref{sample}).
These foreground galaxies could also enhance the probability of 
observing strongly lensed background galaxies.

{\it Zw8197.} --- The hierarchical center is located 8' (667 kpc) W 
of the X-ray center.  The X-ray center lies close to the BCG, but 
there are few/no galaxies around the BCG.  The spatial distribution of 
cluster members is remarkably flat; the BCG lies in a sparse region
of cluster members.  The large magnitude 
gap between the BCG and other cluster members led \citet{santos07} 
to identify Zw8197 as a candidate fossil group.

{\it A2261.} --- This system is near the edge of the DR6 photometric 
footprint.  Although imaging is available through the SDSS Image List 
tool, photometry was not available in DR6 for much of the cluster (this 
situation can arise when a field is taken in poor seeing).   We used the 
Guide Star Catalog to identify likely galaxies in the cluster region and 
the Image List tool to visually inspect these candidates.  We 
experimented with using colors from the Guide Star Catalog 
photometry, but these identifications are less reliable than visual 
classification of the SDSS thumbnails.  DR8 contains photometry for A2261.
We used this photometry to select targets for an 
additional Hectospec pointing.  We increased the sampling density and 
included blue galaxies to examine detailed issues in the strong and weak 
lensing determination of the cluster mass \citep{coe12}.

The D99 binary tree analysis we use for all other clusters 
locates the center of A2261 on a structure 6$\farcm$3 ($\approx 1\Mpc$)  and
400$\kms$ away from the BCG.  This center lies
atop a tight grouping (30$\arcs$) of four bright red cluster members 
(two with $L\sim 10L_*$).
However, the slightly different algorithm for cutting the binary tree
described in \citet{serra11} yields
the cluster center on the BCG.  These two different results are a
consequence of the complex
dynamics of A2261.    
We adopt the center closest to the BCG.  The caustic mass profile is
insensitive to the adopted center.

Two teams have investigated the mass profile of A2261 using gravitational lensing. 
\citet{okabe10} find that A2261 is ``over-concentrated" relative to the expectations 
from numerical simulations. By contrast, \citet{coe12} use HST lensing data to 
measure a concentration closer to the expected value for a massive cluster.  
\citet{coe12} show that lensing estimates of $M_{200}$ can differ by $\sim$25\% 
depending on the assumed cluster geometry (spherical versus triaxial).   
This flexibility allows the lensing mass to agree with various X-ray hydrostatic 
mass estimates that differ by 35\% (and could be affected by non-thermal pressure support).  
However, both the X-ray and lensing mass estimates are larger than 
our estimate of $M_{200}$ \citep[see Figure 12 of][]{coe12}.


\section{\label{ensemble} Ensemble Clusters: The Cluster Mass Profile}

The largest uncertainty in determining dynamical masses of clusters is 
the influence of projection effects.  Stacking clusters together to create an
ensemble cluster can significantly reduce this uncertainty \citep{serra11}.  
We create two sets of ensemble clusters using two methods of grouping 
the clusters.  For all ensemble clusters, we eliminate A1758 and 
MS0906/A750 because they are double clusters (see $\S$\ref{individual}).

First, we divide the clusters into quartiles of $M_{200}$ as determined 
from the caustic mass profile.  We then stack the clusters in physical units 
(positions in kpc, velocities in km s$^{-1}$). 
This method has the disadvantage that it relies on 
the parameters from the caustic mass profiles to assign the clusters into quartiles, 
so the resulting ensemble properties are not completely independent of the 
caustic mass profiles of the individual clusters.

Second, we divide the clusters into quartiles of X-ray luminosity from the 
original ROSAT catalogs.  We then stack the clusters in physical units 
(positions in kpc, velocities in km s$^{-1}$).  This approach avoids any 
use of the individual cluster caustic mass profiles in the stacking 
parameters. 

\begin{figure*}
\plotone{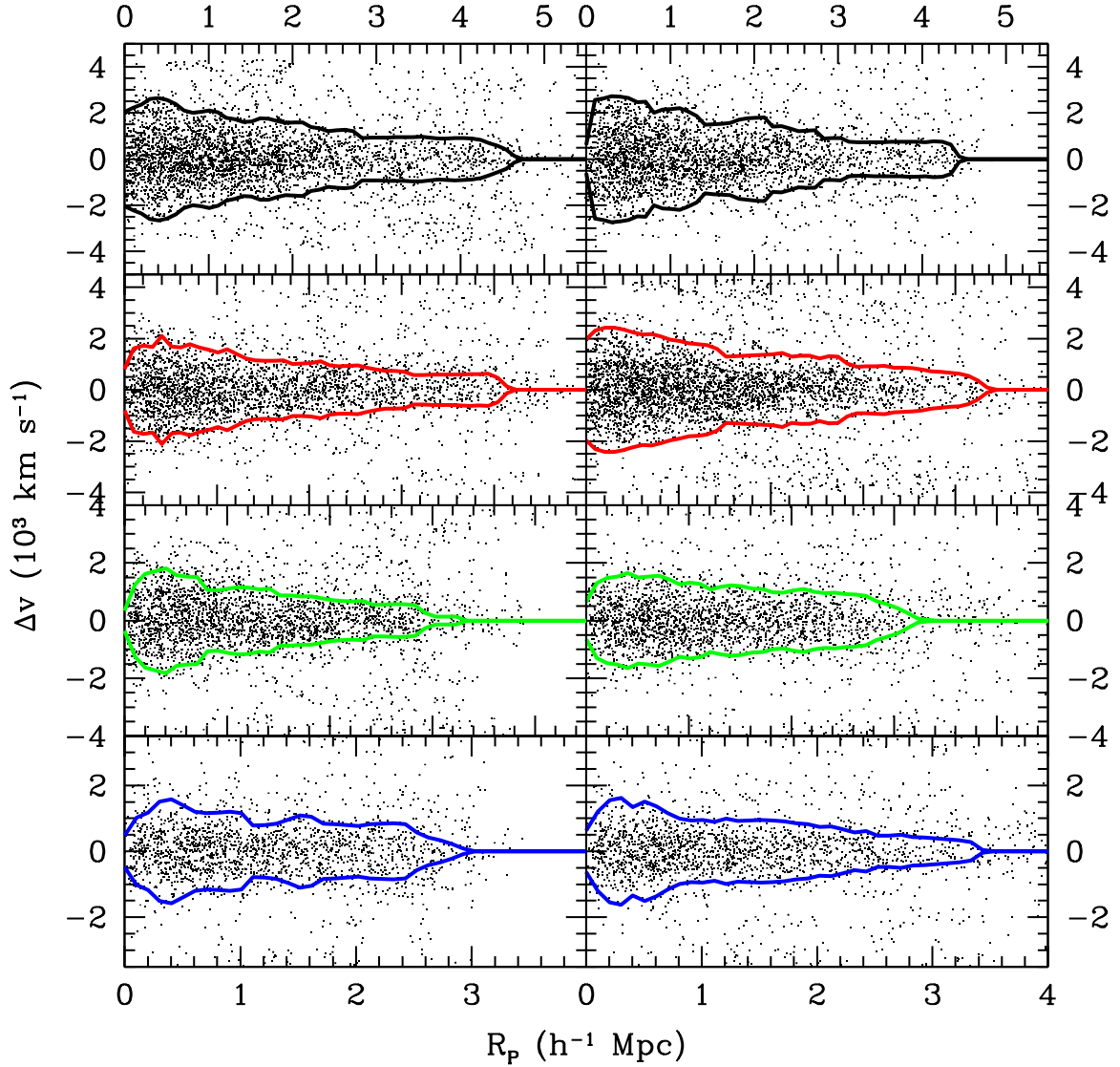}
\caption{\label{quartiles} Redshift versus projected radius in ensemble clusters.  (Left) Quartiles of $L_X$ (Right) Quartiles of $M_{200}$.  Quartiles are shown from top to bottom by decreasing mass/luminosity.  Solid lines indicate the positions of the caustics for the ensemble clusters.}
\end{figure*}

Figure \ref{quartiles} shows the two sets of ensemble clusters.  Both sets show 
a clear distinction between cluster members and field galaxies in all quartiles. 
Table \ref{hecsnfwfits} lists the median and ranges of the quartiles.

\begin{figure*}
\plotone{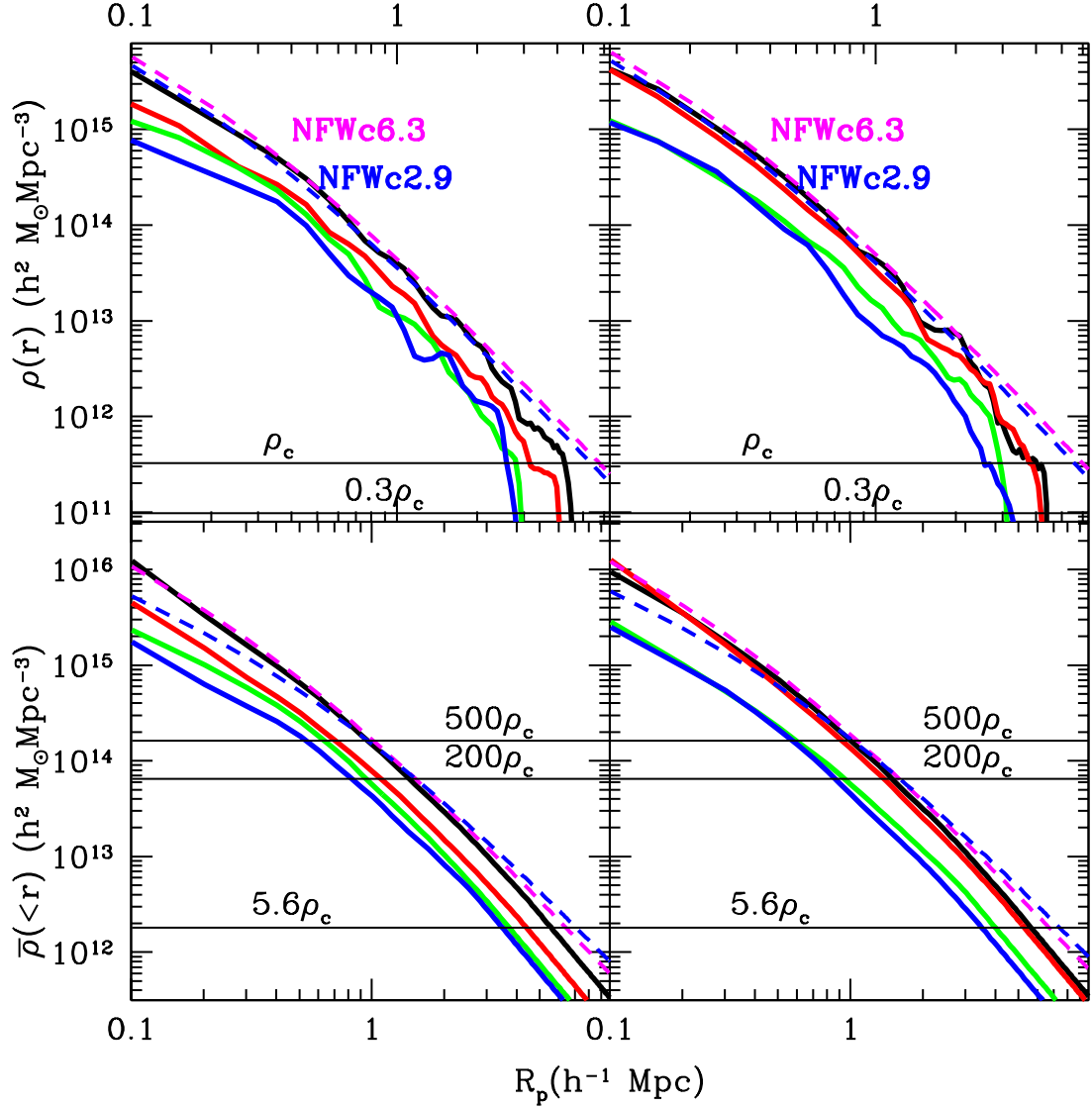}
\caption{\label{rhoquartiles} Density profiles of ensemble clusters.  Black, red, green, and blue 
solid lines show decreasing quartiles of $L_X$ (left panels) or $M_{200}$ (right panels).  
{\it Top panels:} The magenta dashed line shows the NFW profiles that best fit the stacked 
lensing cluster of \citet[][strong-lensing selected]{umetsu11b}.  The blue dashed line shows 
an NFW profile that fits 
the stacked lensing 
cluster of \citet[][X-ray selected]{okabe10b}.  
Horizontal lines indicate the critical density $\rho_c$ and $\bar{\rho}=0.3\rho_c$.  
{\it Bottom panels:} Cumulative density profiles of the ensemble clusters.  Horizontal lines 
indicate enclosed densities of (500,200,5.6)$\rho_{c}$, where $\rho_{c}$ is evaluated at 
$z$=0.16, the median redshift of the HeCS sample. 
}
\end{figure*}

Figure \ref{rhoquartiles} shows the density profiles of the ensemble clusters.  
The caustics trace the mass profiles across nearly four orders of magnitude in density. 
The blue dashed line shows an NFW profile
with a concentration of $c=2.9$.  The low-concentration NFW profile 
is consistent with the inner region of the ensemble cluster profile.  
Concentrations of $c\approx$3-5 are predicted by numerical simulations 
of cluster-sized dark matter haloes \citep[e.g.,][]{bullock01}.  Further, 
\citet{serra11} show that cluster mass profiles in cosmological simulations 
follow the extrapolation of an NFW fit (performed on the inner 1$\Mpc$ 
using the caustic technique) out to 3-4$r_{200}$.  That is, the simulated 
clusters follow NFW profiles well into their infall regions \citep[see also][]{tavio08}.

Some investigators have suggested that the high concentration 
parameters found in cluster lensing profiles indicate problems with 
$\Lambda$CDM cosmology \citep[e.g.,][]{broadhurst08b}.   To ameliorate 
projection effects, \citet{umetsu11b} constructed a stacked cluster and 
determined the mass profile from both strong and weak gravitational 
lensing.  The magenta dashed line in Figure \ref{rhoquartiles} shows an NFW profile with 
concentration $c=r_{200}/r_s=6.28$; this profile is an excellent fit to the 
stacked lensing cluster of \citet[][Umetsu 2012, priv.~comm.]{umetsu11b}, one of the highest-precision 
lensing profile estimates.  Figure \ref{rhoquartiles} also shows the best-fit 
NFW profile ($c$=2.93) from 
the stacked lensing cluster of \citet[][Umetsu 2012, priv.~comm.]{okabe10b} created from X-ray-selected clusters.  
The median mass of the lensing-selected sample is slightly larger than 
the median mass of the highest-$M_{200}$ quartile, but the ranges overlap. 
The high-mass ensemble cluster of \citet{okabe10} has similar $L_X$ and 
$M_{200}$ to our top quartile samples. 
The concentrations of the density profiles of the HeCS ensemble clusters generally lie between
the lensing profile from the X-ray selected clusters and the profile 
from strong-lensing selected clusters (Table \ref{hecsnfwfits}). 

Our results for the ensemble HeCS clusters indicate qualitative agreement 
between the 
simulations and mass profiles determined from the caustic technique.  
Systematic uncertainties could be introduced by our assumption 
of constant $ \mathcal{F}_\beta$, an assumption that overestimates the central masses of simulated clusters
\citep{serra11}.  Another possible concern in measuring $c$ for an ensemble cluster is that
mis-centering could artificially smooth out the central density peak and 
thus artificially lower the measured concentration.  The offset between 
the hierarchical centers we use and the X-ray centers is almost always 
less than 300$\kpc$, but Figure \ref{rhoquartiles} shows that the two 
NFW models are very similar in the range 300-3000$\kpc$, suggesting 
that miscentering could significantly impact concentrations from ensemble
(or individual) caustic mass profiles.  
We will investigate possible systematic 
uncertainties in measuring concentrations with caustics in future work.

The lower panels of Figure \ref{rhoquartiles} show the cumulative density profiles of the 
ensemble clusters.  These profiles determine the values of the 
radii $r_\Delta$. 
For all quartiles, the caustic mass profiles extend 
well beyond $r_{200}$, but not quite to $r_{5.6}$ (the maximum radius of 
particles bound to the cluster in the far future).  This limitation is primarily 
due to the limited radial extent of the HeCS data: a radius of 30$\arcm$
corresponds to 4.2$\Mpc$ at $z=0.2$; the minimum value of 
$r_{5.6}$ for the highest-$L_X$ ensemble cluster is 5.8$\Mpc$.  

\begin{figure*}
\plotone{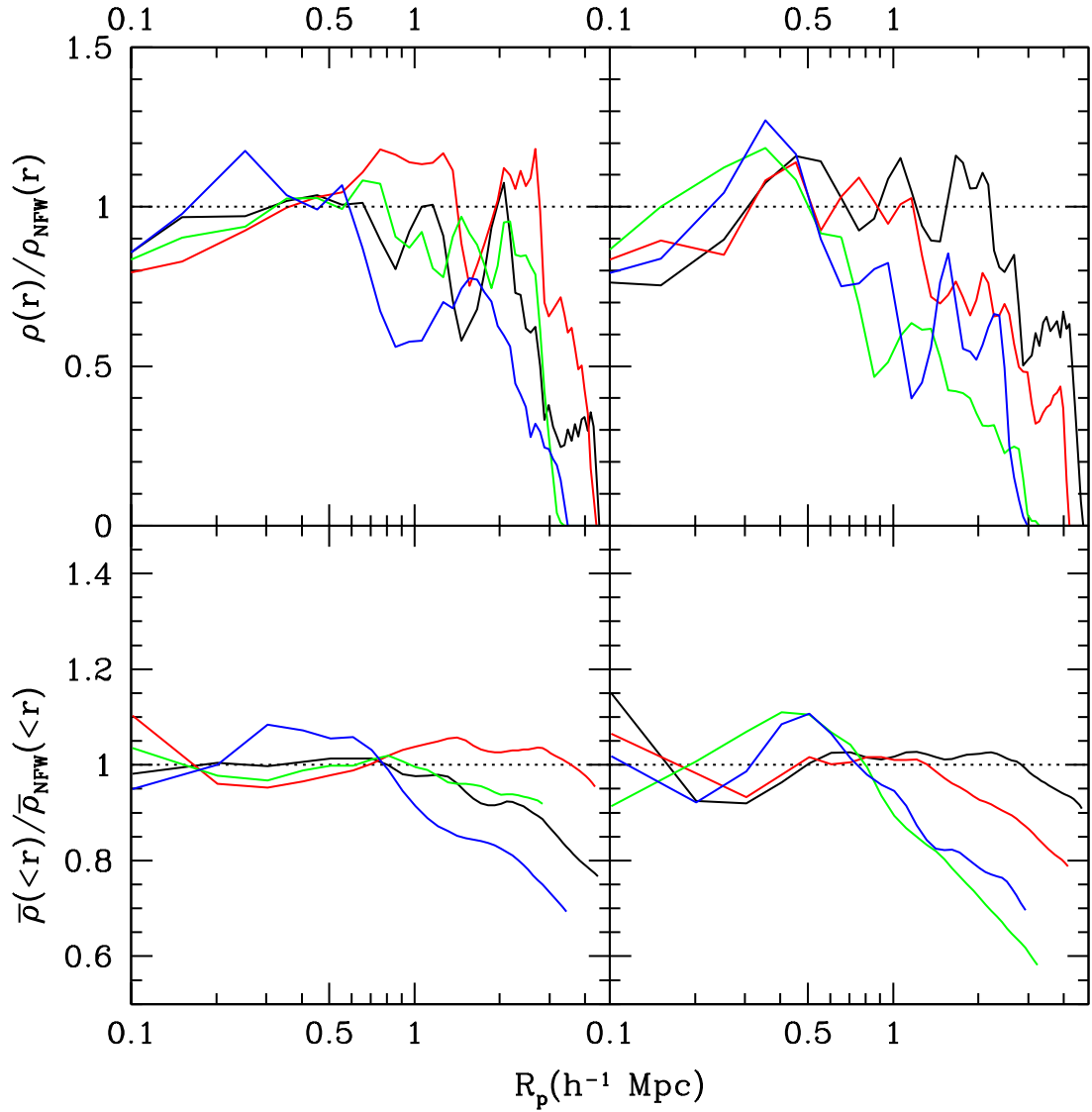}
\caption{\label{rhonfw} Residuals of NFW fits to the density profiles of ensemble clusters from Figure \ref{rhoquartiles}.  Black, red, green, and blue solid lines show decreasing quartiles of $L_X$ (left panels) or $M_{200}$ (right panels).  {\it Top panels} Residuals of the ensemble density profiles from NFW fits.  Note that the NFW fits are performed with the mass profiles rather than the density profiles. {\it Bottom panels} Residuals of cumulative density profiles of the ensemble clusters from NFW fits to the inner 1$\Mpc$  \citep[see][]{serra11}. 
}
\end{figure*}

Figure \ref{rhonfw} shows the residuals from NFW fits to the inner 1$\Mpc$
of the ensemble clusters (fits performed on the caustic mass profile).  \citet{serra11}
show that this procedure yields accurate predictions of their simulated cluster
mass profiles.  The density profiles are noisy but seem to indicate that 
observed densities at large radii are slightly smaller than the densities 
of the extrapolated NFW fits.  Table \ref{hecsnfwfits} lists the best-fit parameters. 
The NFW fits show a positive correlation between concentration and mass 
(or $L_X$) contrary to the expected negative correlation \citep[e.g.,][]{bullock01}. 
Because the concentrations derived from caustic mass profiles may contain 
mass-dependent systematic uncertainties, we caution the reader that the trend
evident in Table \ref{hecsnfwfits} could be dominated by systematic effects.

The lower panels of Figure \ref{rhonfw} 
show that the cumulative density profiles agree within $\sim$20\% of 
the extrapolated NFW profiles far beyond $r_{200}$.  The smallest-$L_X$ quartile 
and the two smallest-$M_{200}$ quartiles have smaller cumulative 
densities than the extrapolated NFW profiles.  We speculate that this deficit 
is a combination of the low-$c$ values of these profile fits ($c\sim$2-3) and 
a dearth of observed galaxies at large radii (Figure \ref{quartiles}) resulting 
from the lower average redshifts of the constituent clusters and the finite 
field-of-view of Hectospec.  
Overall, the agreement between caustic mass profiles and NFW profiles 
extrapolated to large radius is excellent. This dynamical agreement is 
independent support of similar results derived from weak lensing profiles 
extending to large radius \citep{okabe10,umetsu11b}.

\begin{figure*}
\plotone{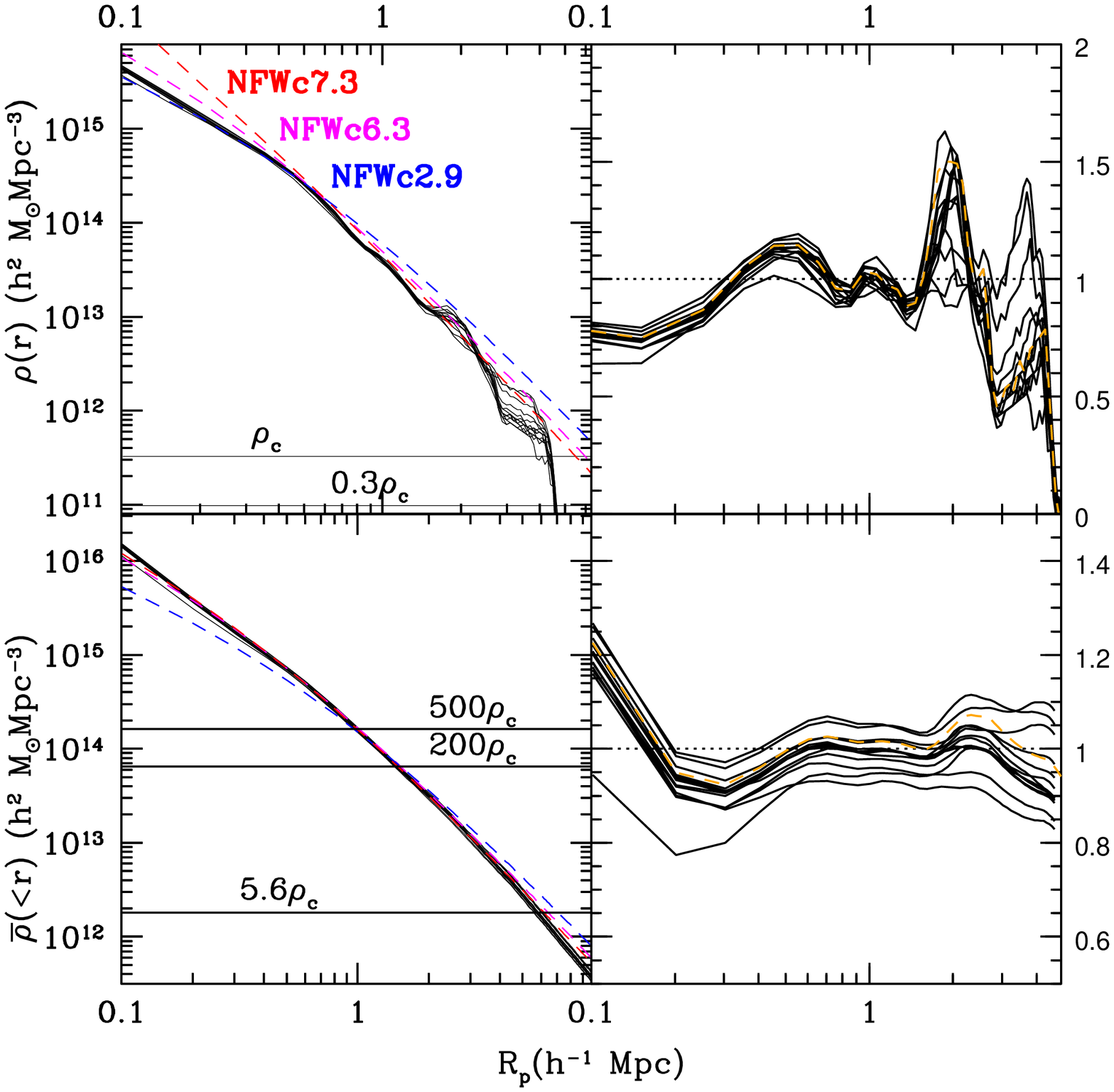}
\caption{\label{rhonfwjack} Density (top left) and enclosed density (bottom left) profiles of the highest-$L_X$ ensemble cluster determined from 13 pseudo-jackknife subsamples.  Magenta, blue, and red dashed lines show NFW profiles with concentrations $c$=6.3, 2.9, and 7.3 respectively (the red line is the best-fit to the enclosed density profile within 1 Mpc).  Each subsample removes a different cluster from the sample.  Right panels show the residuals from the best-fit NFW profile (fit within $1\Mpc$). The orange dashed lines show the profiles determined from all clusters in the high-$L_X$ sample. 
}
\end{figure*}

We use a pseudo-jackknife technique to quantify the statistical uncertainties 
in the ensemble density profiles.  Specifically, we reanalyze the ensemble 
cluster for the highest-$L_X$ quartile in 13 subsamples where we remove 
one cluster from the ensemble for each subsample.  Figure \ref{rhonfwjack}
shows the results of this test; the density profile has statistical uncertainties 
of $\lesssim$10\% inside $r_{200}$ and $\sim$50\% in the range 
(1-4)$r_{200}$.  The enclosed density profile (or equivalently the mass profile) 
from the caustic technique has small ($\lesssim$10\%) statistical uncertainties
at $r<4r_{200}$ (bottom right panel of Figure \ref{rhonfwjack}).

\begin{table}[th] \footnotesize
\begin{center}
\caption{\label{hecsnfwfits} \sc NFW Fits to HeCS Ensemble Mass Profiles}
\begin{tabular}{ccccc}
\tableline
\tableline
\tablewidth{0pt}
Quartile & Median\tablenotemark{a} & Range\tablenotemark{a} & $r_{200}$ & $c_{200}$ \\ 
  &  &  & $\Mpc$ & \\ 
\tableline
$L_X$:1  & 5.06 & 4.14-12.68  & 1.48 & 7.3   \\   
$L_X$:2  & 2.96 & 2.33-3.71  & 1.17 &  4.0  \\   
$L_X$:3  & 1.82 & 1.43-2.10  & 1.07 &   2.5 \\   
$L_X$:4  & 1.11 & 0.62-1.27  & 0.91 &   1.9 \\   
\tableline
$M_{200}:1$  & 7.84 & 4.95-12.40  & 1.63 &  6.0  \\   
$M_{200}:2$  & 3.84  & 3.33-4.58 & 1.44 &  8.8  \\   
$M_{200}:3$  & 2.44  & 1.92-3.10 & 1.02 &  3.0  \\   
$M_{200}:4$  & 1.47  & 0.66-1.86 & 0.95 &  3.2  \\   
Umestu & 9.6 & 8.6-17.2 & -- & 6.28 \\
Okabe & 5.9 & 4.3-10.1 & -- & 2.93 \\
\tableline
\end{tabular}
\tablenote{$L_X/E(z)$ is given in units of $10^{44} h^{-2}\mbox{erg}~\mbox{s}^{-1}$ in the {\em ROSAT} band; $M_{200}$ is in units of $10^{14}h^{-1}M_\odot$. }
\end{center}
\end{table}

\section{\label{galaxyspectra} Properties of Cluster Galaxies}

One goal of HeCS is to study the spectroscopic properties of galaxies 
in clusters and their infall regions.  We defer a complete analysis to 
future work.  Here, we investigate the possible impact of our target
selection algorithm (which favors red-sequence galaxies) on the 
estimates of cluster mass profiles. 

\subsection{\label{redsequence} Red Sequence Galaxies}

\begin{figure}
\plotone{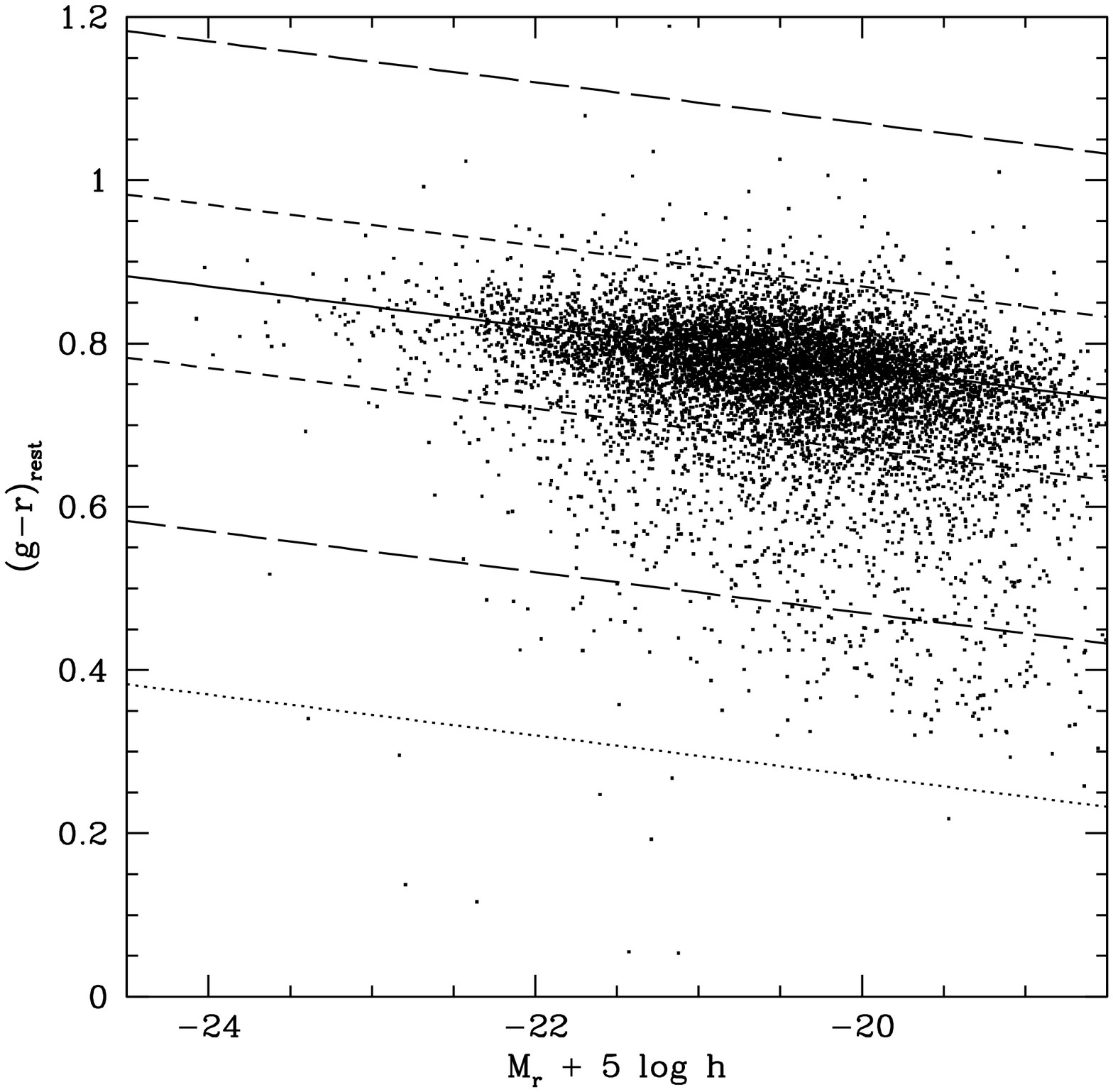}
\caption{\label{hecsgr} Color-magnitude diagram of member galaxies of 
the HeCS clusters including K-corrections.  Long-dashed lines show 0.3 mag 
away from the red sequence, approximately the limits of our priority 
target selection.  Most of the HeCS members are within 0.1 mag of the 
red sequence (short-dashed lines), indicating that the color selection includes the vast 
majority of red-sequence cluster galaxies.  The dotted line shows the color range 
0.2 mag blueward of the red-sequence cut used to select additional targets 
to fill unused fibers.}
\end{figure}

Because we targeted galaxies on the red sequence
(Appendix), many of the cluster members lie on the red
sequence.  Figure \ref{hecsgr} shows color-magnitude diagrams of
cluster members.  We computed K-corrections using the purely 
empirical K-corrections of \citet[][these empirically-based bandpass
corrections naturally include evolutionary corrections]{westra10}.
The red sequence of HeCS members can be described approximately as
\beqn
^{0.0}(g-r) = -0.025(M_r+24)+0.87
\eeqn
(solid line in Figure \ref{hecsgr}).
One striking feature of Figure \ref{hecsgr} is the
small scatter in the red sequence.  The outer lines are offset from the 
red sequence by $\pm$0.3 mag, approximately the color range where 
we assigned highest targeting priority.  The inner lines are offset 
by $\pm$0.1 mag, demonstrating that most cluster members lie within 
this much narrower range of color.  
This result suggests that the
galaxy properties are very similar and that the SDSS photometric
uncertainties are minimal.  Lines with a slope of -0.04 (observed colors) 
or -0.025 (rest-frame colors) provide a
remarkably accurate description of the red sequences.

\begin{figure}
\plotone{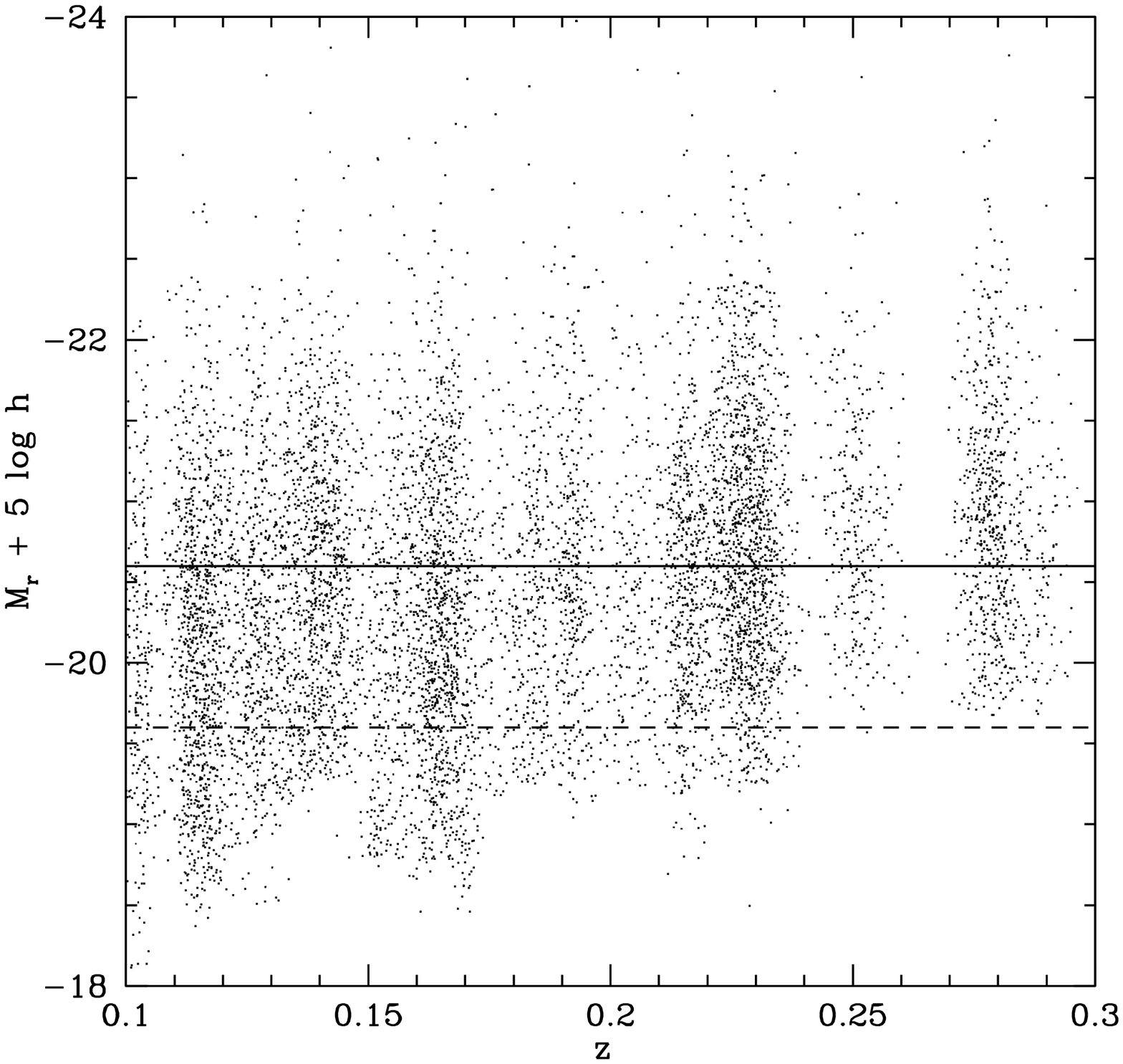}
\caption{\label{hecszrabs} Absolute magnitude versus clustrocentric radius for cluster members in the HeCS  clusters.  Solid and dashed lines indicate absolute magnitudes of $M^*_r$ and $M_r^*+1$ respectively. }
\end{figure}

Figure \ref{hecszrabs} shows the absolute magnitudes of cluster members
versus their redshifts. The solid line show $M_r=-20.60+5\rm{log}h$,
approximately the characteristic magnitude $M_r^*$ of the luminosity 
function of field galaxies in SDSS \citep{blanton03}.  The dashed line 
is one magnitude fainter ($M_r^*+1$), a luminosity limit often adopted 
for describing galaxies \citep[e.g.,][]{tinker05}.  CIRS shows that this limit is
the minimum depth for a cluster sample that yields enough members to identify caustics.
Figure \ref{hecszrabs} shows that all HeCS clusters 
at $z<0.25$ meet this condition.  The remaining clusters are still 
well-sampled because these clusters have larger X-ray luminosities
(and hence more bright member galaxies) than the CIRS clusters 
(Figure \ref{hecslxz}).
 
\subsection{Extremely Red Cluster Galaxies}

Figure \ref{hecsblue} shows that very few cluster members have $g-r$ colors 
significantly redder than the red sequence.  In fact, all of the five cluster 
members in RXJ2129 and A2261 with colors redder than the red-sequence 
cut have incorrect colors due to their proximity to nearby bright stars or 
galaxies.  Using fiber magnitudes, all six have $g-r$ colors below the 
red-sequence cutoff. 

Because the red sequence represents some of the oldest known stellar 
populations, colors redder than the red sequence would most likely arise 
from extreme dust reddening.  Thus, the absence of cluster members with 
colors more than 0.3 mag redder than the red sequence indicates that 
very few cluster members are extremely reddened. 

\subsection{\label{bluegalaxies} Blue Cluster Galaxies: Are They Common? Do They Affect Dynamical Mass Estimates?}

\begin{figure*}
\plotone{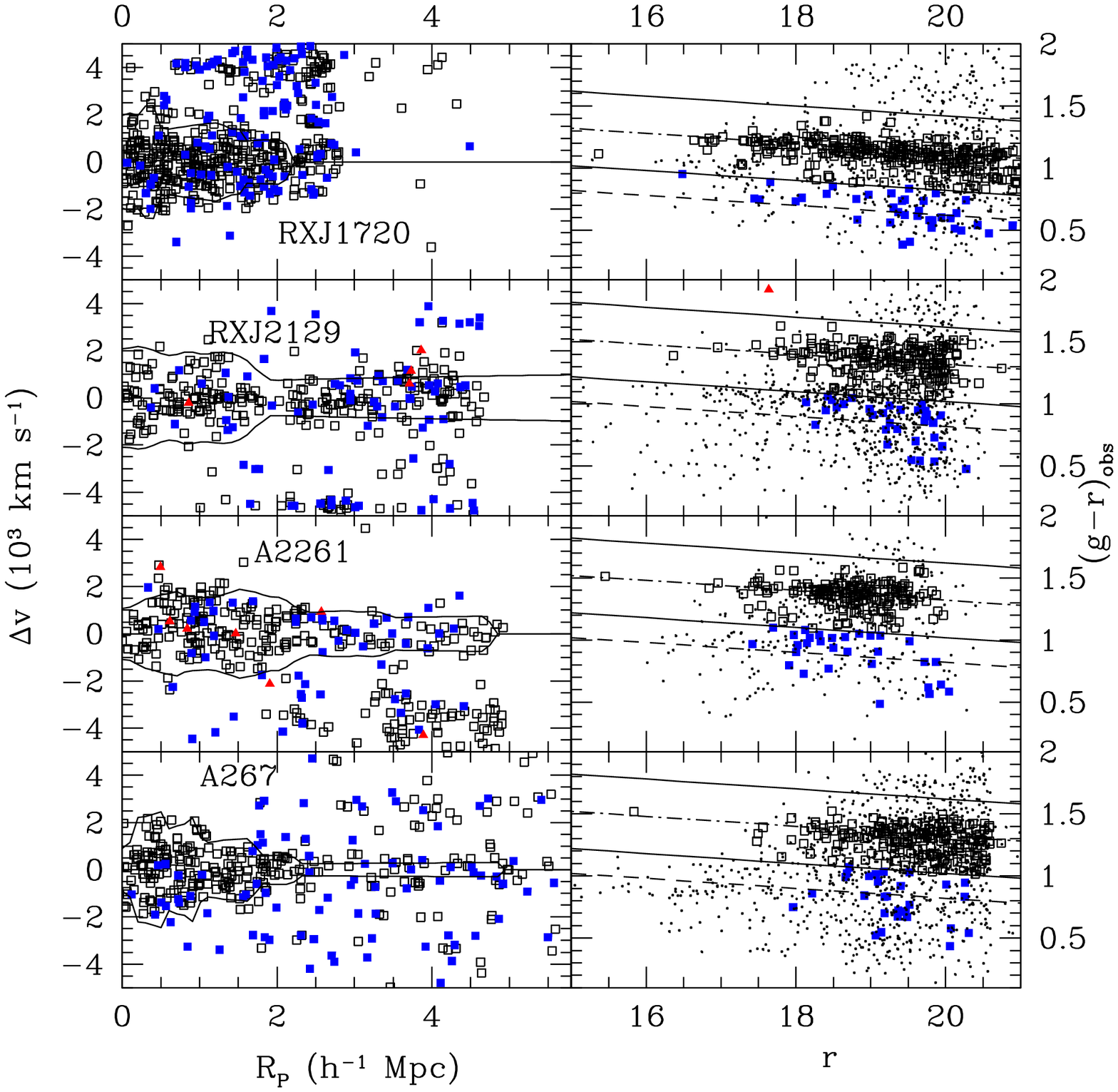}
\caption{\label{hecsblue} (Left) Redshift (rest-frame clustrocentric velocity) versus radius for galaxies in RXJ1720, RXJ2129, A2261 and A267.  For these clusters, spectroscopic redshifts are available for a wide range of colors (see $\S$ 2.3).  Open squares are red sequence galaxies (within 0.3 mag), solid blue squares are blue cloud galaxies, and red triangles are more than 0.3 mag redder than the red sequence. (Right) Color-magnitude diagrams of the four clusters.  Large squares are cluster members (as defined by the caustics) and small dots are non-members. }
\end{figure*}

To test the sensitivity of our dynamical estimates on the red-sequence target selection, 
we observed several Hectospec pointings in the clusters A267, A2261, and RXJ2129 
(all at $z$$\approx$0.23) to sample blue galaxies.  The Hectospec survey 
of RXJ1720 \citep{owers11} also samples both red-sequence and blue galaxies. 
Figure \ref{hecsblue} shows that very few blue galaxies 
are cluster members, consistent with previous studies \citep[e.g.,][]{dressler80,cairnsha}.  
The solid lines show the approximate limits of the HeCS priority target 
selection used for the other HeCS clusters.  The dashed line slightly bluer 
indicates our limit for low-priority targets.  Very few galaxies below the 
solid lines (filled blue squares in the left panels) are cluster members.
Combining the four clusters, 138 of 1053 cluster members (13.1\%) lie blueward
of our red-sequence cut.  A two-sample K-S test indicates that the velocity 
distributions of red and blue galaxies are consistent with being drawn 
from the same parent population.  Further, the velocity dispersion of the 
ensemble cluster (all galaxies) is only 0.8\%  larger than the velocity dispersion
of the red-sequence galaxies.  Limiting the sample to galaxies inside 
$r_{200}$, 38 of 488 members (7.8\%) lie blueward of the red-sequence cut; 
including non-red-sequence galaxies increases the velocity 
dispersion by 0.3\%.  We thus conclude that targeting red-sequence
galaxies produces no significant bias in our estimates of dynamical masses or 
velocity dispersions.  Previous claims of velocity segregation often relied on much 
smaller samples where membership classification may have been less robust.  
\citet{serra12} discusses the completeness of galaxy samples identified using 
the caustic technique in simulations.  They conclude that the caustics identify 
most of the members and include few interlopers. 

\citet{mahajan11} recently studied the dependence of cluster galaxy 
properties on their projected velocity relative to the cluster center.  
They bin galaxies both by projected radius and by velocity offset (e.g., galaxies 
with $|\Delta v|$=(0-1)$\sigma_p$, 1-2$\sigma_p$, 2-3$\sigma_p$). At fixed 
projected radius, their Figure 3 shows little dependence of the color 
distribution on velocity offset, suggesting that the velocity distribution does 
not depend dramatically on galaxy color, consistent with our results. 

Our Hectospec (and SDSS) redshifts show that some cluster members are
``blue cloud'' galaxies with colors indicating recent star formation
(Figure \ref{hecsblue}).  The fraction of ``blue cloud'' galaxies seems
to increase with increasing absolute magnitude, similar to field
galaxies \citep[e.g.,][]{blanton03b}.  This conclusion could be tested with more
extensive spectroscopic sampling of faint galaxies that lie blueward
of the HeCS red sequence cuts.

\section{Conclusions}

We present \nczhecsnew redshifts from the Hectospec Cluster Survey (HeCS).  HeCS is a 
MMT/Hectospec spectroscopic survey of X-ray-selected clusters contained 
in the imaging footprint of SDSS DR6.  Our redshifts confirm that infall patterns
known as ``caustics" are clearly present in X-ray clusters at moderate redshift. 
Combined with SDSS data, we define a sample of \nczmem cluster members.

We use the infall patterns to compute mass profiles for the clusters
extending in many cases to the turnaround radius of the cluster.  In 
numerical simulations of a $\Lambda$CDM universe, the 
mass within the radius $r_{5.6}$ (the average density inside $r_{5.6}$ 
is $5.6\rho_{c}(z)$) is approximately equal to the 
ultimate mass of the cluster at late times.  These simulations predict 
that this ultimate halo mass is $\approx$2.2$M_{200}$.  The HeCS 
mass profiles provide an observational estimate of $M_{5.6}=(1.99\pm0.11)M_{200}$, in 
excellent agreement with the predictions. 

The caustic technique enables a unique measure of the large-scale behavior of
cluster mass profiles. The striking agreement with the theoretical predictions
\citep{nl02,busha05,dunner06} is an interesting new test of 
models for the growth of structure in the $\Lambda$CDM cosmology.

The ensemble clusters we construct here average over projection effects.
The density profiles of these ensemble clusters closely resemble NFW 
profiles out to radii of at least $2r_{200}$.  The NFW concentrations of our ensemble 
clusters are similar to or slightly larger than the predictions of simulations.  
Systematic uncertainties from projection effects and from determining the cluster 
prevent us from confirming or rejecting the
``over-concentration" problem of strong-lensing clusters \citep[e.g.,][]{broadhurst08}.
Future work will test the precision and accuracy of measurements of NFW 
concentrations from caustic mass profiles and compare these measurements
to predictions from numerical simulations of dark matter haloes.  

The CAIRNS and CIRS projects demonstrated that caustic patterns are 
present in nearly all rich, X-ray luminous galaxy clusters.  Here we show 
that these patterns are also prominent in clusters at 
$z$=0.1-0.3. The ensemble clusters we construct enhance the 
visibility of these patterns.  

The HeCS clusters show 
a relation between $L_X$ and $\sigma_p$ (we will investigate the 
specific relation in future work); we show that outliers in this relation 
can be attributed to contamination from X-ray point sources (A689) 
or strong cooling cores (RXJ1504).
The scaling relation $M_{200}-\sigma_p$ between virial masses and 
line-of-sight velocity dispersions is in excellent agreement with the 
scaling relation of dark matter particles in simulated clusters \citep{evrard07}.
%
Other projects planned with HeCS are investigations of cluster scaling  
relations between galaxy dynamics and other mass probes \citep[richness, 
$Y_X$, $Y_{SZ}$,$M_{lens}$, see][]{hecsysz}, a determination of the 
mass function of clusters \citep[e.g.,][]{rines08,vikhlinin09a,mantz08,henry09}
and a detailed study of the photometric and spectroscopic properties 
of cluster galaxies.  A companion paper \citep{geller12b} compares 
caustic mass profiles to those determined from weak gravitational lensing, 
the only other mass estimator that applies to the non-virialized infall regions 
surrounding clusters.

\acknowledgements

KR was funded by a Cottrell College Science Award from the Research 
Corporation.  AD acknowledges partial support from the INFN grant PD51
and the PRIN-MIUR-2008 grant ``Matter-antimatter asymmetry, dark matter
and dark energy in the LHC Era''.  Research support for MJG and MJK is provided by the Smithsonian Institution.
We thank Yu-Ying Zhang, Keiichi Umetsu, Rien van de Weygaert, 
and the anonymous referee 
for helpful comments and suggestions.

Funding for
the Sloan Digital Sky Survey (SDSS) has been provided by the Alfred
P. Sloan Foundation, the Participating Institutions, the National
Aeronautics and Space Administration, the National Science Foundation,
the U.S. Department of Energy, the Japanese Monbukagakusho, and the
Max Planck Society. The SDSS Web site is http://www.sdss.org/.  The
SDSS is managed by the Astrophysical Research Consortium (ARC) for the
Participating Institutions. 
The Participating Institutions are The
University of Chicago, Fermilab, the Institute for Advanced Study, the
Japan Participation Group, The Johns Hopkins University, the Korean
Scientist Group, Los Alamos National Laboratory, the
Max-Planck-Institute for Astronomy (MPIA), the Max-Planck-Institute
for Astrophysics (MPA), New Mexico State University, University of
Pittsburgh, University of Portsmouth, Princeton University, the United
States Naval Observatory, and the University of Washington.

{\it Facility:} \facility{MMT:Hectospec}

\bibliographystyle{apj}
\bibliography{rines}

\appendix

\section{\label{hectospectra} Spectroscopic Target Selection}

We selected targets for spectroscopy based on photometry from SDSS DR6.   
First, we extracted galaxy catalogs within 32$\arcm$ from the SDSS SQL server\footnote{http://cas.sdss.org/astrodr6/en/tools/search/sql.asp}.  
For each cluster, we identify the red sequence from a color-magnitude 
diagram using $g-r$ colors and $r$-band apparent magnitudes.  
Following guidance from the SDSS web pages\footnote{http://www.sdss.org/dr6/algorithms/photometry.html}, 
we adopt the composite model magnitudes for our apparent magnitudes and 
colors.  Composite model magnitudes are a linear combination of the best-fit 
exponential and deVaucouleurs model profiles.  Composite model magnitudes 
should have higher signal-to-noise than Petrosian magnitudes, especially 
for relatively faint galaxies such as those studied here. 

To identify the red sequence, we used 
a fixed slope of -0.04 ${\rm mag}~{\rm mag}^{-1}$ for all clusters and chose the 
intercept based on visual inspection of the color-magnitude diagram\footnote{For 8 clusters 
observed in 2007 (A2111, A2187, A2219, A2259, A2396, A2631, A2645, and RXJ2129), we used a simple $g-r$ color cut rather than a red-sequence cut.  
This small sampling difference has no significant impact on our results.}.
To eliminate some stars and artifacts (e.g., portions of diffraction spikes 
from bright stars), we restrict the sample to galaxies with $r$-band surface 
brightness brighter than 22.9 ${\rm mag}~{\rm arcsec}^{-2}$ (we later inspect 
objects with fainter surface brightnesses to include some galaxies 
eliminated by this cut).  

The Hectospec fiber assignment software {\it xfitfibs}\footnote{\url{https://www.cfa.harvard.edu/~john/xfitfibs/}} allows the user 
to assign rankings to targets.  If we ignored the spatial positions of our targets,
fiber collisions would prevent many objects in the central portion of the field 
(usually the center of the cluster) from being observed.  We therefore assigned 
highest priority to galaxies brighter than a cluster-dependent limiting magnitude 
within $\pm$0.3 magnitudes of the red sequence and within 2.4$\arcm$ of the 
X-ray center.  We give second priority to red galaxies near the center and up to one 
magnitude fainter.  Priorities 
are then assigned in annuli (outer radii 7$\farcm$5 and 15$\arcm$) according to 
apparent magnitude. In general, our target catalogs contained approximately 
twice as many targets as the number of fibers available.  Most 
bright targets in the inner 15$\arcm$ are assigned fibers, as are many fainter 
targets in this region.  Unassigned targets are typically in the outer parts of the 
Hectospec field, with a modest bias against targets in local overdensities due to 
fiber collisions.
Later analysis shows
that the $\pm$0.3 mag band around the red-sequence is significantly 
larger than the actual thickness of the red-sequence (see $\S$\ref{redsequence}); thus, the precise choice 
of red-sequence intercept has essentially no impact on our results.  
Because 
cluster members are centrally concentrated, we found that prioritizing 
central objects is critical to obtain reasonably uniform sampling as a function 
of radius. 

Because 
fiber placement constraints would sometimes prevent fibers from being 
deployed, we add galaxies up to 0.2 magnitude bluer than our red-sequence
cutoff to the target list.  We use the SDSS Image List tool\footnote{http://cas.sdss.org/dr6/en/tools/chart/list.asp}
to visually inspect all targets prior to observation.  This visual inspection 
identifies artifacts and bright stars.  We added a small number of targets 
that are not identified as separate objects by the DR6 photometric pipeline 
(these targets are usually merged with a nearby star or galaxy).

We observed two Hectospec pointings per cluster for clusters at $z$$>$0.15 
and one pointing per cluster for clusters at $z$=$0.10-0.15$.  We adopted this 
strategy because SDSS redshifts of Main Sample galaxies at $r$$<$17.77 provide 
many members for the lower-redshift clusters (although not enough to enable 
a full caustic analysis).  Also, the lower-redshift clusters have generally smaller 
X-ray luminosities (because HeCS is flux-limited) and thus have smaller expected masses.  
Combined with the larger angular size of these clusters, one Hectospec pointing 
combined with SDSS redshifts is generally sufficient to avoid sampling bias. 
For clusters at $z$$>$0.15, we chose apparent magnitude limits for each cluster to select 900-1300 candidate 
targets for two Hectospec pointings (potentially up to 600 targets).  The bright magnitude limit for these clusters 
is 18.5$<r<$20.0, corresponding to $M^*_r+1$ or fainter for most clusters (Figure \ref{hecszrabs}).  
For clusters at $z=$0.10-0.15, we chose apparent magnitude limits for each cluster to select 500-600 candidate 
targets for a single Hectospec pointing (potentially up to 300 targets).  The bright magnitude limit for these clusters 
is 17.8$<r<$18.8, again corresponding to $M^*_r+1$ or fainter for most clusters (Figure \ref{hecszrabs}).

\begin{figure*}
\plotone{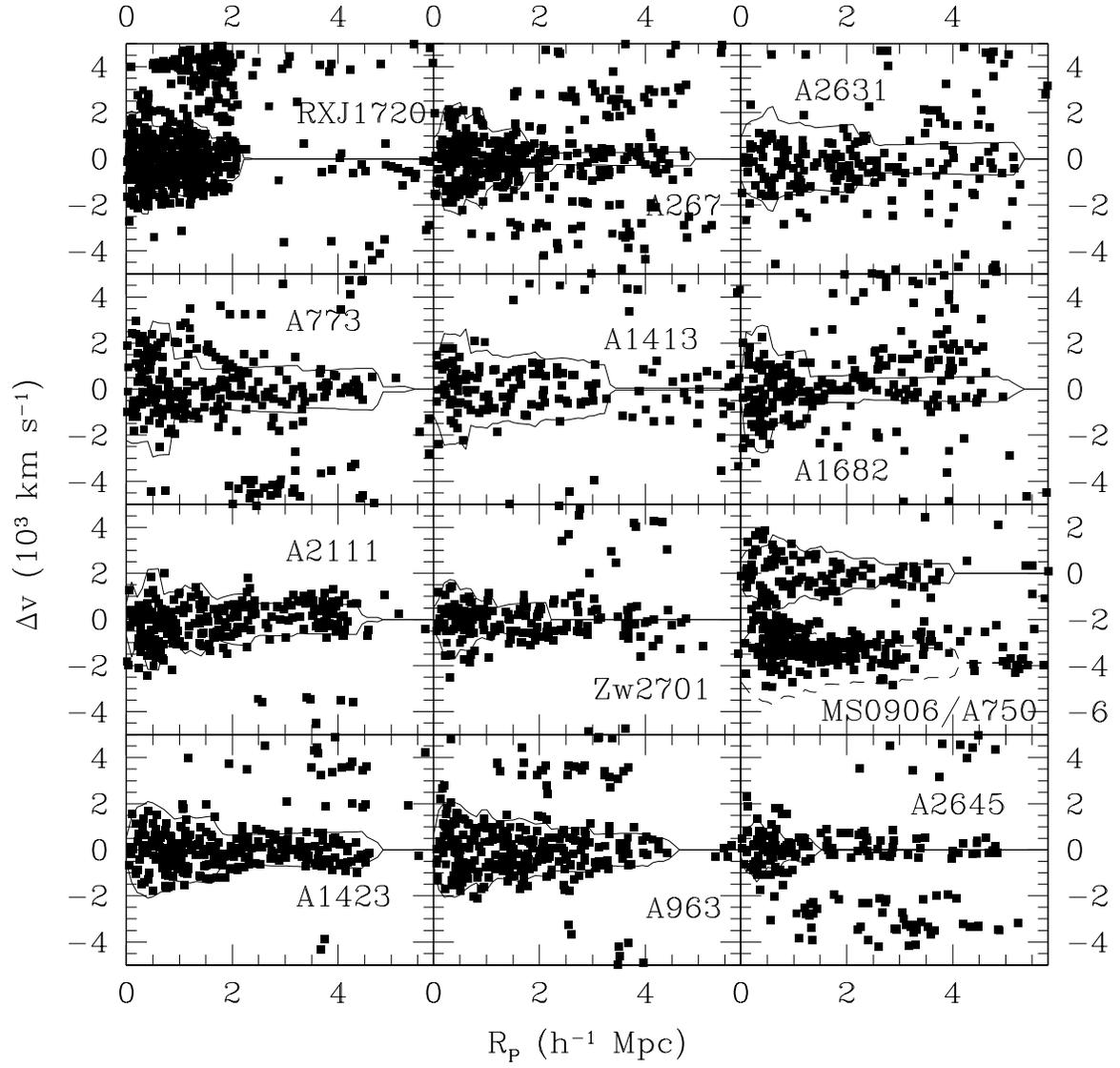}
\caption{ Redshift (rest-frame clustrocentric velocity) versus projected radius for galaxies around
HeCS clusters.  The caustic
pattern is evident as the trumpet-shaped regions with high density.
The solid lines indicate our estimate of the location of the caustics
in each cluster.  Clusters are ordered left-to-right and top-to-bottom
by decreasing X-ray luminosity. }
\end{figure*}

\begin{figure*}
\plotone{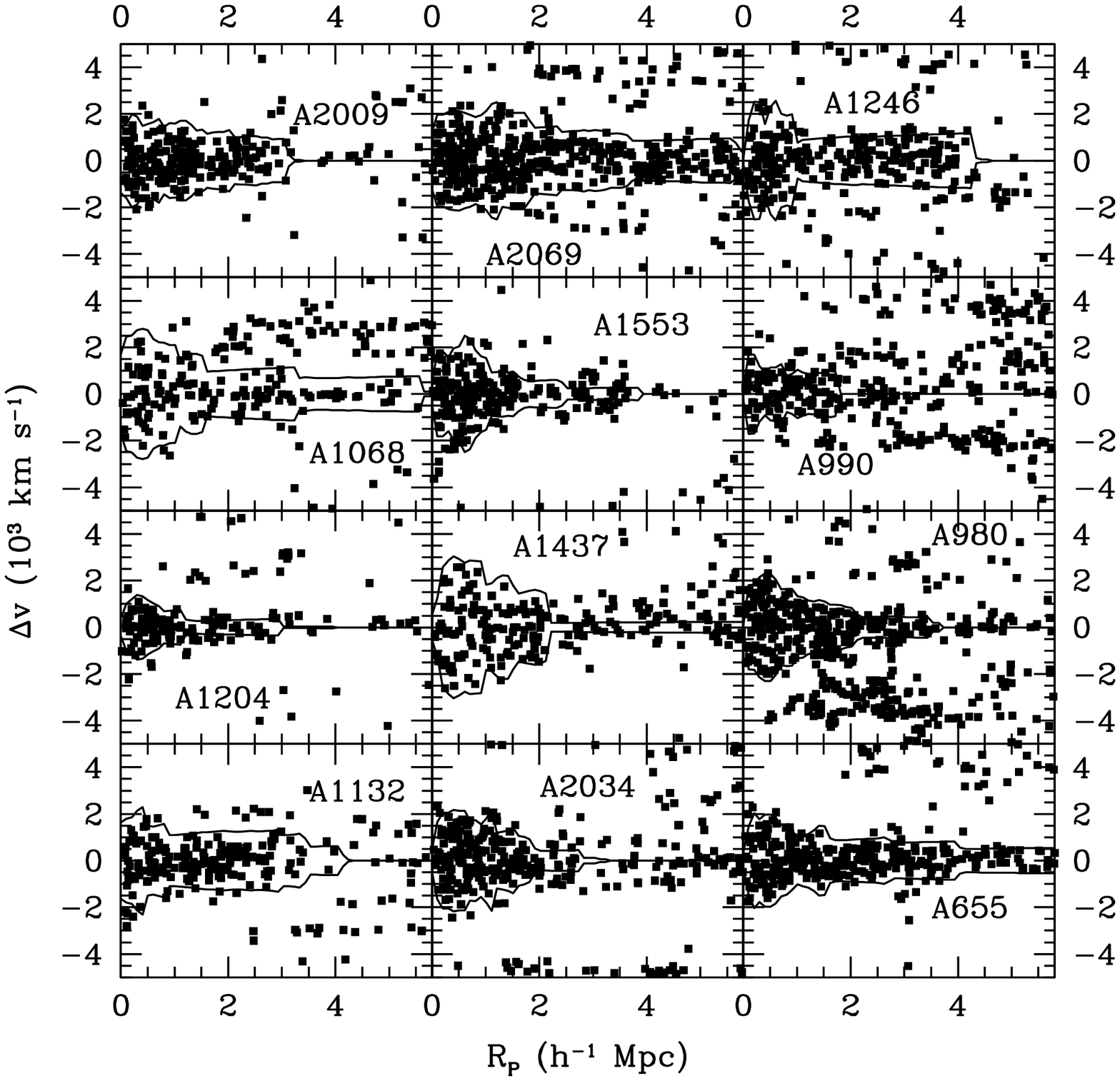}
\caption{ Redshift (rest-frame clustrocentric velocity) versus projected radius for galaxies around
HeCS clusters.  The caustic
pattern is evident as the trumpet-shaped regions with high density.
The solid lines indicate our estimate of the location of the caustics
in each cluster.  Clusters are ordered left-to-right and top-to-bottom
by decreasing X-ray luminosity. }
\end{figure*}

\begin{figure*}
\plotone{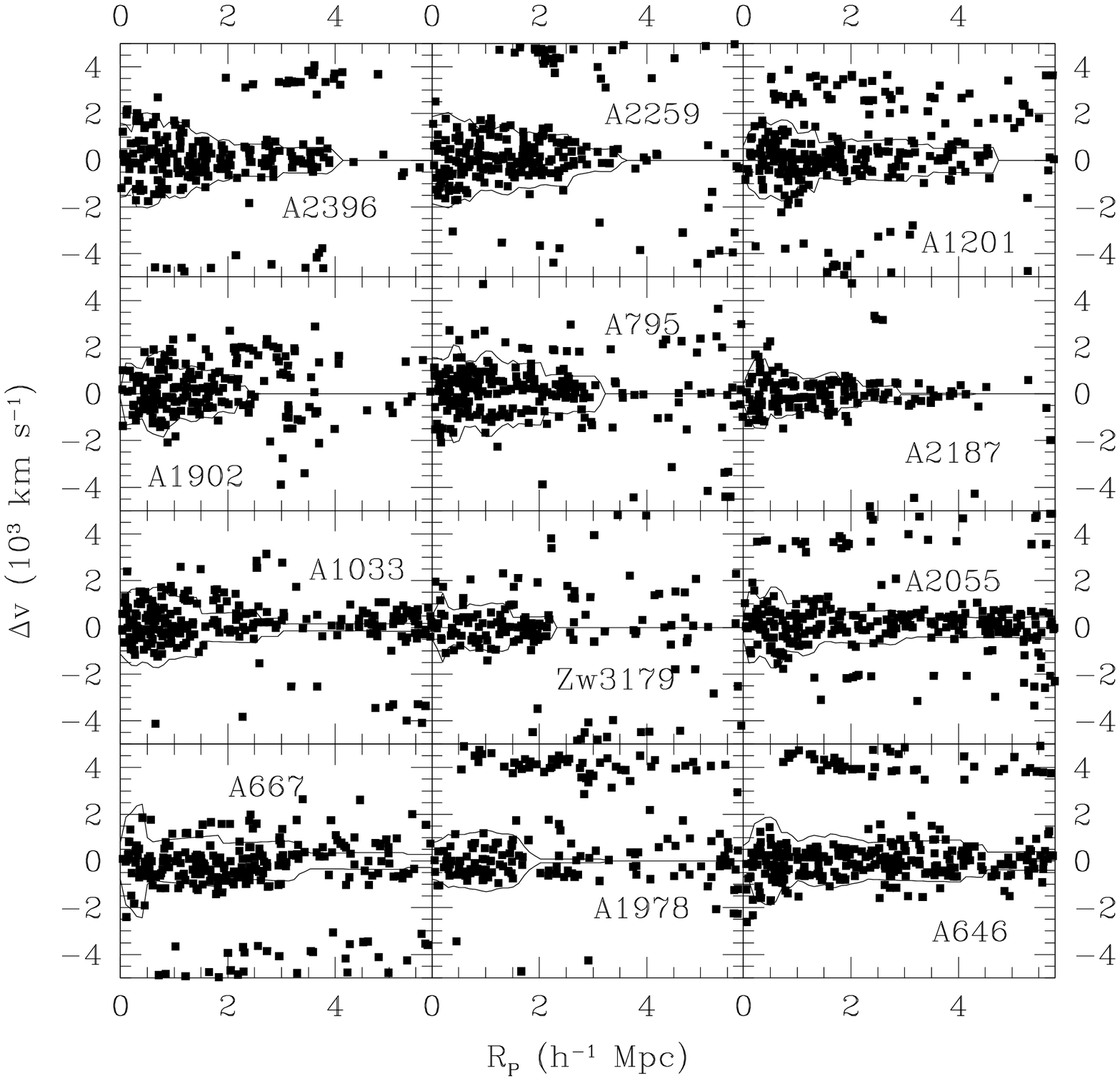}
\caption{ Redshift (rest-frame clustrocentric velocity) versus projected radius for galaxies around
HeCS clusters.  The caustic
pattern is evident as the trumpet-shaped regions with high density.
The solid lines indicate our estimate of the location of the caustics
in each cluster.  Clusters are ordered left-to-right and top-to-bottom
by decreasing X-ray luminosity. }
\end{figure*}

\begin{figure*}
\plotone{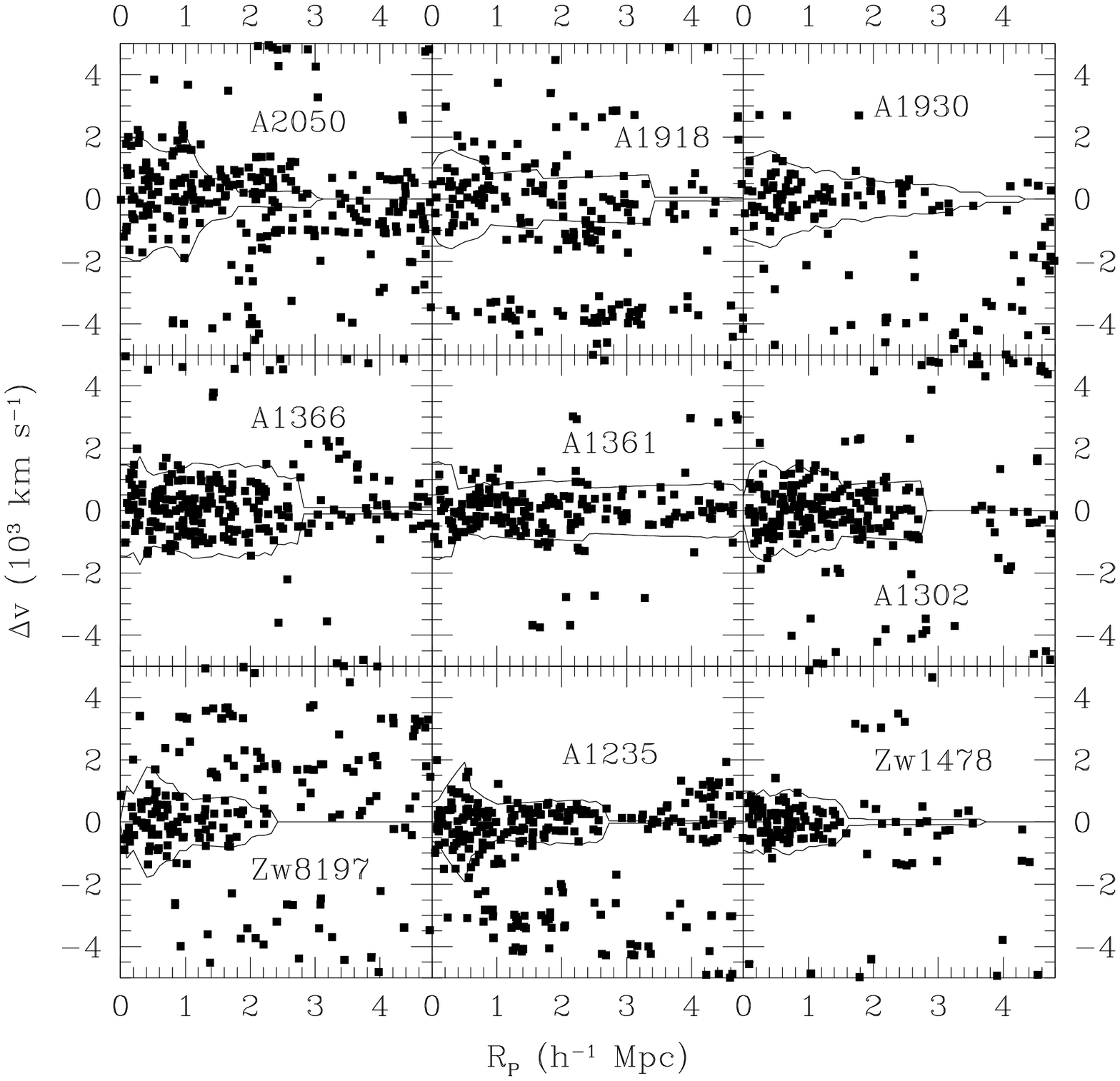}
\caption{ Redshift (rest-frame clustrocentric velocity) versus projected radius for galaxies around
HeCS clusters.  The caustic
pattern is evident as the trumpet-shaped regions with high density.
The solid lines indicate our estimate of the location of the caustics
in each cluster.  Clusters are ordered left-to-right and top-to-bottom
by decreasing X-ray luminosity. }
\end{figure*}

\begin{figure*}
\plotone{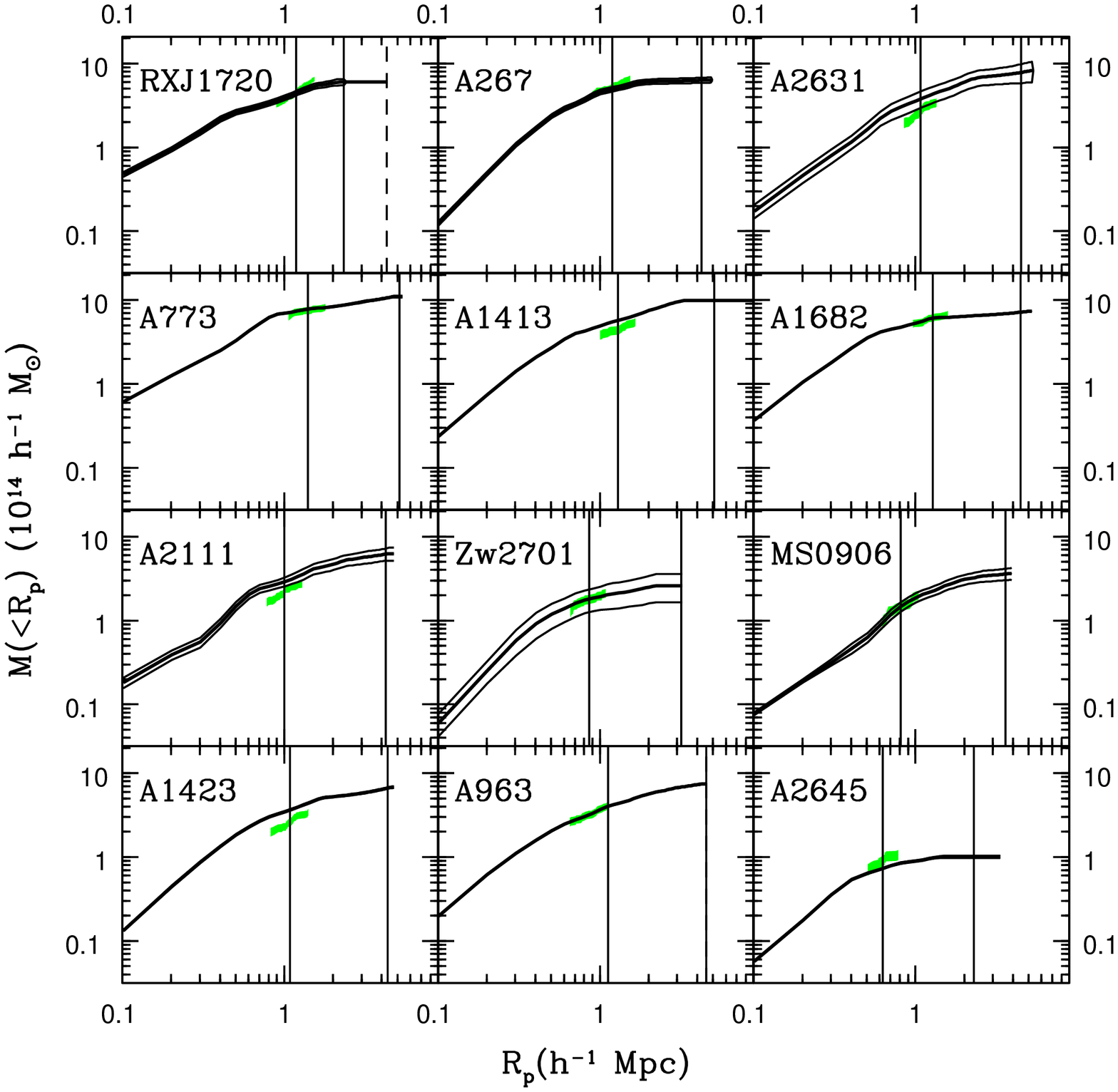}
\caption{\label{allhecsm2} Caustic mass profiles for the HeCS  clusters.
The thick solid lines show the caustic mass profiles and the thin
lines show the 1$\sigma$ uncertainties in the mass profiles. The inner vertical 
solid line in each panel shows the radius $r_{200}$.  The next vertical 
line shows the smaller of  $r_{5.6}$ (the limit of bound structure) and 
$r_{max}$ (the maximum radius where the caustics are detected). 
For clusters with $r_{max}<r_{5.6}$, a dashed vertical line shows a 
lower limit on $r_{5.6}$ assuming no mass is present outside $r_{max}$.
Green shaded regions indicate the virial mass profile in the range
(0.75-1.3)$r_{200}$ (approximately from $r_{500}$ to $r_{100}$). 
}
\end{figure*}

\begin{figure*}
\plotone{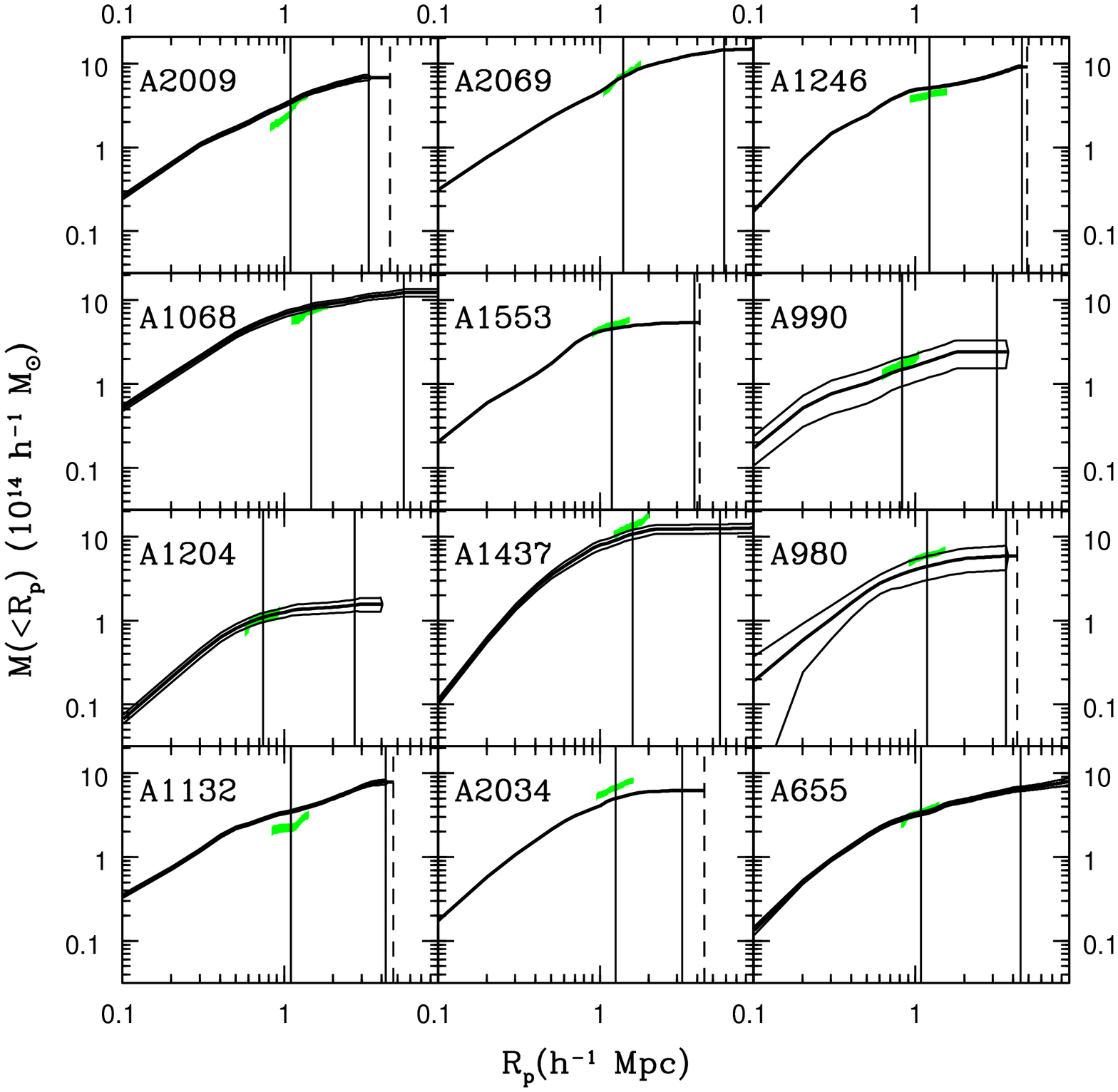}
\caption{\label{allhecsm3} Caustic mass profiles for the HeCS  clusters.
The thick solid lines show the caustic mass profiles and the thin
lines show the 1$\sigma$ uncertainties in the mass profiles. The inner vertical 
solid line in each panel shows the radius $r_{200}$.  The next vertical 
line shows the smaller of  $r_{5.6}$ (the limit of bound structure) and 
$r_{max}$ (the maximum radius where the caustics are detected). 
For clusters with $r_{max}<r_{5.6}$, a dashed vertical line shows a 
lower limit on $r_{5.6}$ assuming no mass is present outside $r_{max}$.
Green shaded regions indicate the virial mass profile in the range
(0.75-1.3)$r_{200}$ (approximately from $r_{500}$ to $r_{100}$). 
}
\end{figure*}

\begin{figure*}
\plotone{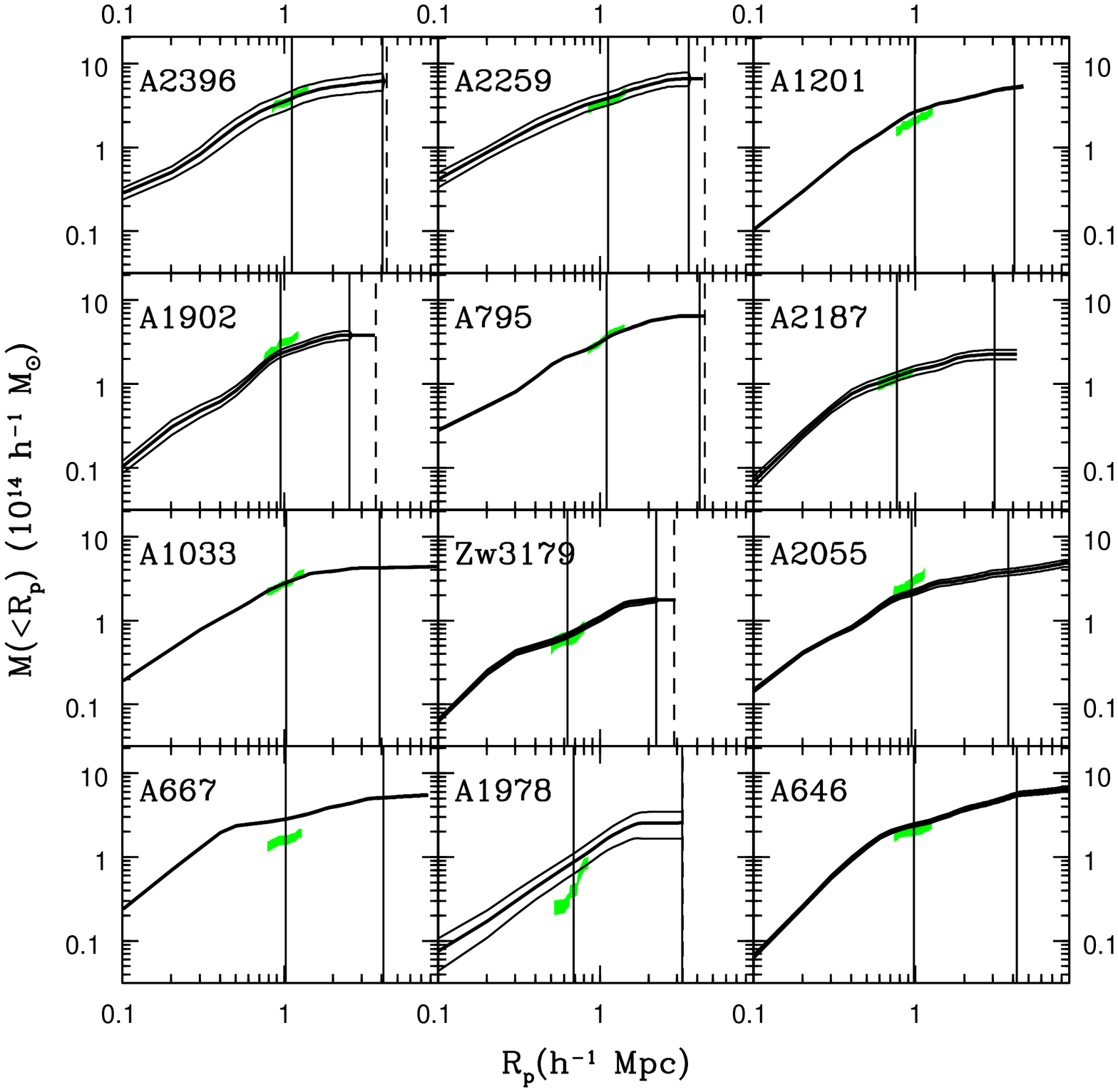}
\caption{\label{allhecsm4} Caustic mass profiles for the HeCS  clusters.
The thick solid lines show the caustic mass profiles and the thin
lines show the 1$\sigma$ uncertainties in the mass profiles. The inner vertical 
solid line in each panel shows the radius $r_{200}$.  The next vertical 
line shows the smaller of  $r_{5.6}$ (the limit of bound structure) and 
$r_{max}$ (the maximum radius where the caustics are detected). 
For clusters with $r_{max}<r_{5.6}$, a dashed vertical line shows a 
lower limit on $r_{5.6}$ assuming no mass is present outside $r_{max}$.
Green shaded regions indicate the virial mass profile in the range
(0.75-1.3)$r_{200}$ (approximately from $r_{500}$ to $r_{100}$). 
}
\end{figure*}

\begin{figure*}
\plotone{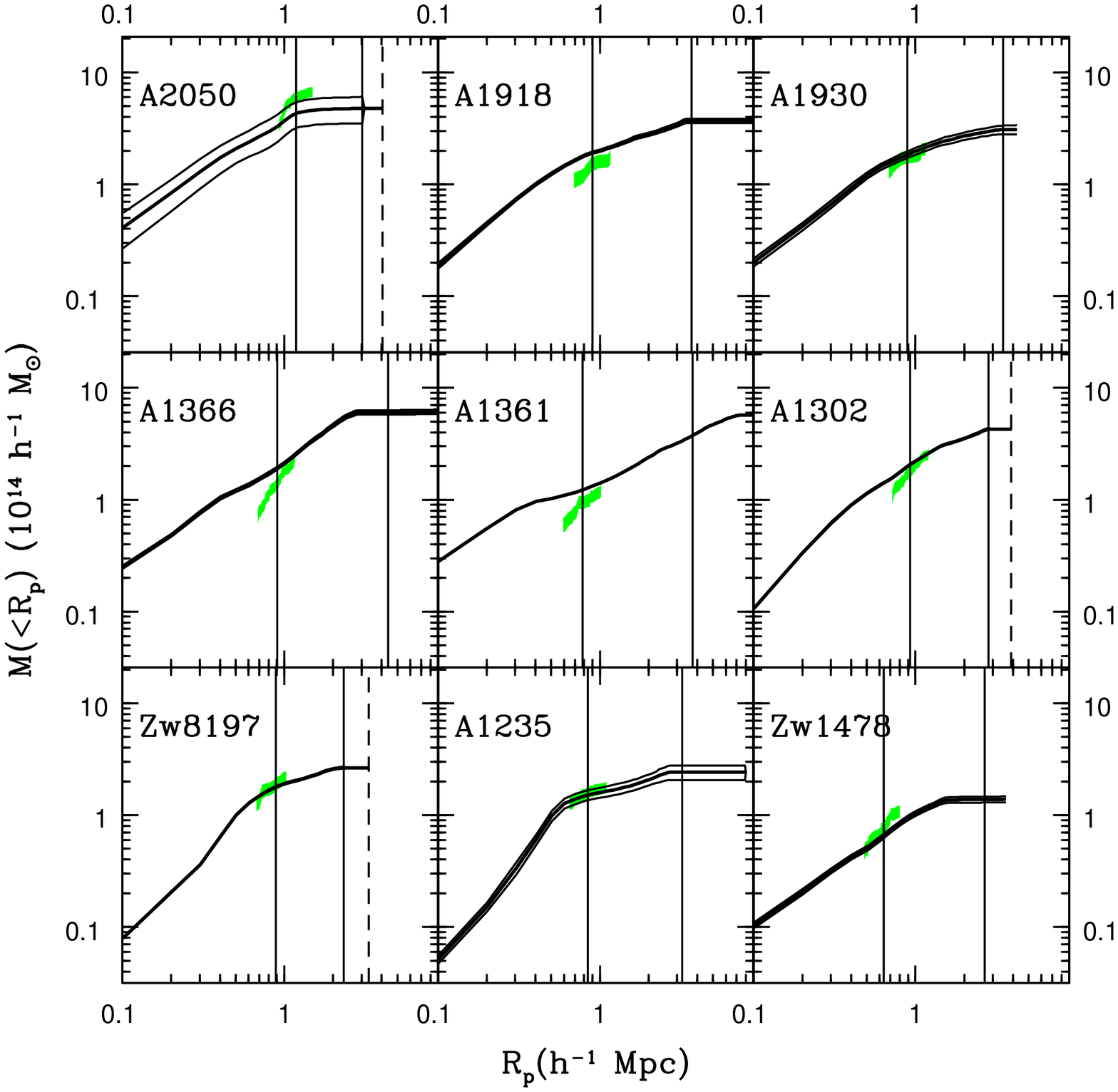}
\caption{\label{allhecsm5} Caustic mass profiles for the HeCS  clusters.
The thick solid lines show the caustic mass profiles and the thin
lines show the 1$\sigma$ uncertainties in the mass profiles. The inner vertical 
solid line in each panel shows the radius $r_{200}$.  The next vertical 
line shows the smaller of  $r_{5.6}$ (the limit of bound structure) and 
$r_{max}$ (the maximum radius where the caustics are detected). 
For clusters with $r_{max}<r_{5.6}$, a dashed vertical line shows a 
lower limit on $r_{5.6}$ assuming no mass is present outside $r_{max}$.
Green shaded regions indicate the virial mass profile in the range
(0.75-1.3)$r_{200}$ (approximately from $r_{500}$ to $r_{100}$). 
}
\end{figure*}

\end{document}